\documentclass[11pt]{article}
\usepackage{epsfig}
\usepackage{amsmath}
\usepackage{latexsym}
\usepackage{graphicx}
\usepackage{amsfonts}
\usepackage{amssymb}

\newtheorem{definition}{Definition}
\newtheorem{example}{Example}

\newtheorem{lemma}{Lemma}

\newtheorem{proposition}{Proposition}

\newenvironment{proof}[1][Proof]{\noindent\textbf{#1.}
}{\hspace*{\fill}\ \rule{0.5em}{0.5em} \vspace{2ex}}

\DeclareMathOperator{\dom}{dom}
\DeclareMathOperator{\probop}{P}

\DeclareMathOperator{\expecop}{E}

\DeclareMathOperator{\varop}{Var}

\newcommand{\card}[1]{\left|{#1}\right|}

\DeclareMathOperator{\expdistrib}{Exp}

\DeclareMathOperator{\poissondistrib}{Poisson}

\DeclareMathOperator{\bigO}{O}
\DeclareMathOperator{\intrinsic}{I}
\DeclareMathOperator{\insertion}{I}
\DeclareMathOperator{\deletion}{D}
\DeclareMathOperator{\modification}{M}

\newcommand{\intdspace}{\hspace{0.15em}}

\begin{document}

\title{Managing Periodically Updated Data in Relational Databases: \\A Stochastic Modeling Approach}
\author{Avigdor Gal\thanks{ Department of Management Science and Information Systems,
Rutgers University, Piscataway, NJ 08854 USA, phone: (732) 445-3245, fax:
(732) 445-6329, e-mail: \texttt{avigal@rci.rutgers.edu} }
\and Jonathan Eckstein\thanks{ Department of Management Science and Information
Systems\ and RUTCOR, Rutgers University, Piscataway, NJ 08854 USA, phone:
(732) 445-0510, fax: (732) 445-6329, e-mail:
\texttt{jeckstei@rutcor.rutgers.edu}}}
\date{}
\maketitle
\begin{abstract}
Recent trends in information management involve the periodic transcription of
data onto secondary devices in a networked environment, and the proper
scheduling of these transcriptions is critical for efficient data management.
To assist in the scheduling process, we are interested in modeling
\emph{data obsolescence}, that is, the
reduction of consistency over time between a relation and its replica. The
modeling is based on techniques from the field of stochastic processes, and
provides several stochastic models for content evolution in the base relations
of a database, taking referential integrity constraints into account. These
models are general enough to accommodate most of the common scenarios in
databases, including batch insertions and life spans both with and without
memory. As an initial ``proof of concept'' of 
the applicability of our approach, we validate the insertion portion of
our model framework
via experiments with real data feeds. We also discuss a set of
transcription protocols which make use of the proposed stochastic model.
\end{abstract}

\section{Introduction and motivation}

Recent developments in information management involve the transcription of
data onto secondary devices in a networked environment, \emph{e.g.},
materialized views in data warehouses and search engines, and replicas in
pervasive systems. Data transcription influences the way databases define and
maintain consistency. In particular, the networked environment may require
periodic (rather than continuous) synchronization between the database and
secondary copies, either due to paucity of resources (\emph{e.g.}, low
bandwidth or limited night windows) or to the transient characteristics of the
connection. Hence, the consistency of the information in secondary copies,
with respect to the transcription origin, varies over time and depends on the
rate of change of the base data and on the frequency of synchronization.

Systematic approaches to the proper scheduling of transcriptions necessarily
involve optimizing a trade-off between the cost of transcribing fresh
information versus the cost of using obsolescent data. To do so, one must
quantify, at least in probabilistic terms, this latter cost, which we call
\emph{obsolescence cost} \cite{GAL99c}. This paper aims to provide a
comprehensive stochastic framework for quantifying time-dependent data
obsolescence in replicas. Suppose we are given a relation $R$, a start time
$s\in\Re$, and some later time $f>s$. We denote the extension of a relation
$R$ at time $t\in\Re$ by $R(t)$. Starting from a known extension $R(s)$, we
are interested in making probabilistic predictions about the contents of the
later extension $R(f)$. We also suggest a cost model schema to quantify the
difference between $R(s)$ and $R(f)$. Such tools assist in optimizing the
synchronization process, as demonstrated in this paper. Our approach is based
on techniques from the field of stochastic processes, and provides several
stochastic models for content evolution in a relational database, taking
referential integrity constraints into account. In particular, we make use of
compound nonhomogeneous Poisson models and Markov chains; see for example
\cite{ROSS80,ROSS95,TK94}. We use Poisson processes to model the behavior of
tuples entering and departing relations, allowing (nonhomogeneous) time-varying behavior ---
\emph{e.g.}, more intensive activity during work hours, and less intensive
activity after hours and on weekends --- as well as compound (bulk) 
insertions, that is,
the simultaneous arrival of several tuples. We use Markov chains in a
general modeling approach for attribute modifications, allowing the assignment
of a new value to an attribute in a tuple to depend on its current value. The
approach is general enough to accommodate most of the common scenarios in
databases, including batch insertions and memoryless, as well as time
dependent, life spans.

As motivation, consider the following two examples:

\begin{example}
[Query optimization]Query optimization relies heavily on estimating the
cardinality and value distribution of relations in a database. If these
statistics are outdated and inaccurate, even the best query optimizer may
formulate poor execution plans. Typically, statistics are updated only
periodically, usually at the discretion of the database administrator, using
utilities such as DB2's RUNSTATS. Although some research has been devoted to
speeding up statistics collection through sampling and wavelet
approximations~\cite{HAAS95,MATIAS98}, periodic updates are unavoidable in
very large databases such as IBM's Net.Commerce \cite{SHURETY98}, an
e-business software package with roughly one hundred relations, or an SAP
application, which has more than 8,000 relations and 9,000 indices. Collection
of statistics becomes an even more acute problem in database federations
\cite{SHETH90}, where the federation members do not always ``volunteer'' their
statistics \cite{RASCHID2001} (or their cost models for that matter
\cite{ROTH96}), and are unwilling to burden their resources with frequent
statistics collection.

In current practice, cardinality or histogram data recorded at time $s$ are
used unchanged until the next full analysis of the database at some later time
$s^{\prime}>s$. If a query optimization must be performed at some time
$f\in(s,s^{\prime})$, the optimizer simply uses the statistics gathered at
time $s$, since the time spent recomputing them may overwhelm any benefits of
the query optimization. As an alternative, we suggest using a probabilistic
estimate of the necessary statistics at time $f$. Use of these techniques
might make it possible to increase the interval between statistics-gathering
scans, as will be discussed in Example~\ref{ex:qopt2}.\hspace*{\fill}$\Box$
\end{example}

\begin{example}
[Replication management in distributed databases]\label{ex:replica} We now
consider replication management in a distributed database. Since fully
synchronous replication management, in which a user is guaranteed access to
the most current data, comes at a significant computational cost, most
commercial distributed database providers have adopted asynchronous
replication management. That is, updates to relation replicas are performed
after the original transaction has committed, in accordance with the workload
of the machine on which the secondary copy is stored. Asynchronous
replicas are also
very common in Web applications such as search engines, where Web crawlers
sample Web sites periodically, and in \emph{pervasive systems }(\emph{e.g.},
Microsoft's Mobile Information
Server\footnote{http://www.microsoft.com/servers/miserver/} and Caf\'{e}
Central\footnote{http://www.comalex.com/central.htm}). In a pervasive system,
a server serves many different users, each with her own unpredictable
connectivity schedule and dynamically changing device capabilities. Our
modeling techniques would allow client devices to reduce the rate at which
they poll the server, saving both server resources and network
bandwidth. We demonstrate the usefulness of stochastic modeling in
this setting in Sections~\ref{sec:condepup} 
and~\ref{sec:exwebcraw}.\hspace*{\fill}$\Box$
\end{example}

The novelty of this paper is in developing a formal framework for modeling
content evolution in relational databases. The problem of content evolution
with respect to materialized views (which may be regarded as a complex form of
data transcription) in databases has already been recognized. For example, in
\cite{ABITEBOUL99}, the incompleteness of data in views was noted as being a
``dynamic notion since data may be constantly added/removed from the view.''
Yet, we believe that there has been no prior formal modeling of the evolution
process.\footnote{Other research efforts involve probabilistic database
systems (\emph{e.g.}, \cite{LAKSHMANAN97}), but this work is concerned with
uncertainty in the stored data, rather than data evolution.} Related research
involves the containment property of a materialized view with respect to its
base data: a few of the many references in this area include
\cite{YANG87,CHAUDHURI95,LEVY95a,ABITEBOUL98,GRUMBACH00}. However, the
temporal aspects of content evolution have not been systematically addressed
in this work. In \cite{ABITEBOUL98}, for example, the containment
relationships between a materialized view $I$ and the ``true'' query result
$\mathcal{V}(D)$, taken from a database $D$, can be either $I=\mathcal{V}(D)$
or $I\subseteq\mathcal{V}(D)$. The latter relationship represents a situation
where the materialized view stores only a partial subset of the query result.
However, taking content evolution into account, it is also possible that
$I\supset\mathcal{V}(D)$, if tuples may be deleted from $\mathcal{V}(D)$ and
$I$ is periodically updated. Moreover, modifications to the base data may
result in both $I\not \subseteq\mathcal{V}(D)$ and $I\not \supseteq
\mathcal{V}(D)$.

Refresh policies for materialized views have been previously discussed in the
literature (\emph{e.g.}, \cite{LINDSAY86} and \cite{COLBY96}). Typically,
materialized views are refreshed immediately upon updates to the base data, at
query time (as in \cite{COLBY96}), or using snapshot databases (as in
\cite{LINDSAY86}). The latter approach can produce obsolescent materialized
views. A combination of all three approaches appears in \cite{COLBY97}. Our
methodology differs in that we do not assume an \emph{a priori} association of
a materialized view with a refresh policy, but instead design policies based
on their transcription and obsolescence costs.

A preliminary attempt to describe the time dependency of updates in the
context of Web management was given in \cite{CHO00}, which suggests a simple
homogeneous Poisson process to model the updating of Web pages. We suggest
instead a nonhomogeneous compound Poisson model, which is far more flexible,
and yet still tractable. In addition, the work in~\cite{CHO00} supposes that
transcriptions are performed at uniform time intervals, mainly because
``crawlers cannot guess the best time to visit each site.'' We show in this
paper that our model of content evolution gives rise to other, better
transcription policies.

In \cite{OLSTON00}, a trade-off mechanism was suggested to decide between the
use of a cache or recomputation from base data by using range data, computed
at the source. In this framework, an update is ``pushed'' to a replication
site whenever updated data falls outside a predetermined interval, or whenever
a query requires current data. The former requires the client and the server
to be in touch continuously, in case the server needs to track down the
client, which is not always realistic (either because the server does not
provide such services, or because the overhead for such services undermines
the cost-effectiveness of the client). The latter requirement puts the burden
of deciding whether to refresh the data on the client, without providing it
with any model for the evolution of the base data. We attempt to fill this gap
by providing a stochastic model for content evolution, which allows a client
to make judicious requests for current data. Other work in related areas
(\emph{e.g.}, \cite{ALONSO90,CAREY91a,DELIS98}) has considered
various alternatives for pushing updated data from a server to a cache on
the client side. Lazy replica-update policies using replication graphs have
also been discussed in, for example, \cite{ANDERSON98}. This work, however,
does not take the data obsolescence into account, and is primarily concerned
with transaction throughput and timely updates, subject to network constraints.

As with models in general, our model is an idealized representation of a
process. To be useful, we wish to make predictions based on tractable
analytical calculations, rather than detailed, computationally intensive
simulations. Therefore, we restrict our modeling to some of the more basic
tools of applied probability theory, specifically those relating to Poisson
processes and Markov chains. Texts such as~\cite{ROSS80,TK94} contain the
necessary reference material on Markov chains and Poisson processes, and
specifically on nonhomogeneous Poisson processes. Poisson processes can model
a world where data updates are independent from one another. In databases with
widely distributed access, \emph{e.g.}, Web interfacing databases, such an
independence assumption seems plausible, as was verified in \cite{CHO00}.

The rest of the paper is organized as follows: Section \ref{sec:preliminaries}
introduces some basic notation. Section~\ref{sec:estcard}
provides a content evolution model for insertions and deletions, while
Section~\ref{sec:modif} discusses data modifications. We
shall introduce preliminary results of fitting the insertion model parameters
to real data feeds in Section \ref{sec:verify}. A cost model
and transcription policies that utilize it follow in Section \ref{costmodel},
highlighting the practical impact of the model. Conclusions and topics for
further research are provided in Section \ref{sec:conclusion}.

\subsection{Notational preliminaries}

\label{sec:preliminaries} In what follows, we denote the set of attributes and
relations in the database by $\mathcal{B}$ and $\mathcal{R}$, respectively.
Each $R\in\mathcal{R}$ consists of a set of attributes $\mathcal{A}%
(R)\subseteq\mathcal{B}$, and also has a \emph{primary key} $\mathcal{K}(R)$,
which is a nonempty subset of $\mathcal{A}(R)$. Each attribute $A\in
\mathcal{B}$ has a \emph{domain} $\dom
A$, which we assume to be a finite set, and for any subset of attributes
$\mathcal{A}=\left\{  A_{1},A_{2},...,A_{k}\right\}  $, we let $\dom
{\mathcal{A}}=\dom A_{1}\times\dom A_{2}\times...\times\dom
A_{k}$ denote the compound domain of $\mathcal{A}$. We denote by $r.A(t)$ the
value of attribute $A$ in tuple $r$ at time $t$, and similarly use
$r.\mathcal{A}(t)$ for the value of a compound attribute. For a given time
$t$, subset of attributes $\mathcal{A}\subseteq\mathcal{A}(R)$, and value
$v=\langle v_{1},v_{2},...,v_{k}\rangle\in\dom{\mathcal{A}}$, we define
$R_{\mathcal{A},v}(t)=\left\{  r\in R(t)\;\left|  \;\;(r.A_{1}(t)=v_{1}%
)\wedge(r.A_{2}(t)=v_{2})\wedge\ldots\wedge(r.A_{k}(t)=v_{k})\right.
\right\}  $. We also define $\hat{R}_{\mathcal{A}}(t)$ to be the
\emph{histogram} of values of $\mathcal{A}$ at time $t$, that is, for each
value $v\in\dom\mathcal{A}$, $\hat{R}_{\mathcal{A}}(t)$ associates a
nonnegative integer $\hat{R}_{\mathcal{A},v}(t)$, which is the cardinality of
$R_{\mathcal{A},v}(t)$.\footnote{This vector can be computed exactly and
efficiently using indices. Alternatively, in the absence of an index for a
given attribute, statistical methods (such as ``probabilistic'' counting
\cite{WHANG90}, sampling-based estimators \cite{HAAS95}, and wavelets
\cite{MATIAS98}) can be applied.}  This notation, and well as other
symbols used throughout the paper, 
are also summarized in Table \ref{tab:listssym}.

\renewcommand{\arraystretch}{1.3}

\begin{table}[t] \centering
{\scriptsize
\begin{tabular}
[c]{|l|p{4.5in}|}\hline
$s,f$ & Points in time\\\hline
$R,S\in\mathcal{R}; R(t); \card{R(s)}$ & Relations; 
$R$'s extension at time $t$; its cardinality at time $s$.\\
$A\in\mathcal{B}; \mathcal{A} \subseteq \mathcal{B}; \dom A;
\dom\mathcal{A}$ & 
Attribute; compound attribute; domain of attribute;
domain of compound attribute \\
$\mathcal{A}(R) \subseteq \mathcal{B}; 
\mathcal{K}(R); 
\mathcal{C}(R)$ & 
Attributes of $R$; primary key of $R$; modifiable attributes of $R$ \\
$r; r.A(t); r.\mathcal{A}(t)$ & 
Tuple; value of attribute $A$ in $r$ at $t$; 
value of compound attribute $\mathcal{A}$ in $r$ at $t$ \\
$v\in\dom\mathcal{A}; 
R_{\mathcal{A},v}(t); \hat{R}_{\mathcal{A}}(t)$ &
Value; set of tuples with $r.A(t)=v$; histogram of
$\mathcal{A}$\\
$b(r); d(r)$ & Insertion time of $r$; deletion time of $r$\\
$\mathcal{N}\subset\mathcal{B}$ & Set of numeric attributes \\
$G; G(R)$ & 
Dependency multigraph; dependency sub-multigraph generated by $R$\\
$\lambda_{R}(t)\hspace{0.15em};\Lambda_{R}(s,f);B_{R}(s,f)$ & 
Insertion rate (intensity); 
expected number of insertion events during $(s,f]$;
number of insertions during $(s,f]$ \\
$\expdistrib_{s}(\phi(\cdot));L_{R,s};L_{R,s}^{\intrinsic}$ & 
Nonhomogeneous exponential distribution; 
interarrival time;
remaining life span
\\\hline
$\Delta_{R,i}^{+}; \Delta_{i}^{-}$ & 
Number of tuples for insertion event $i$;
number of tuples for deletion  event $i$ \\
$\mu_{R}(t); M_{R}(s,f)$ & 
Deletion rate (intensity); expected number of deletion events
\\\hline
$w(r,S)$ & 
Number of tuples in $S$ forcing deletion of $r$ via referential
integrity \\
$W(R,S,t)$ & 
Random variable of $w(r,S)$ over uniform selection of $r\in R$\\
$W(R,t)$ &
Vector of $W(R,S,t)$ over $S\in G(R)$
\\\hline
$p_{R}(s,f)$ & 
Probability that a tuple in $R$ at time $s$ survives through $f$ \\
$\hat{p}_{R}(t,f)$ &
Survival probability through $f$ for tuple inserted at $t$
\\\hline
$\expecop_{r\in R(s)}\!\left[  {\cdot}\right]  $ & 
Expectation over uniform random
selection of tuples $r\in R(s)$ \\\hline
$X_{R}(s,f)$ & 
Number of tuples inserted into $R$ during $(s,f]$ \\
$Y_{R}(s,f)$ &
Number of tuples in $R(s)$ surviving through $f$\\
$Y_{R}^{+}(s,f); Y_{R}^{-}(s,f)$ &
Surviving tuples that were modified;
surviving tuples that were not modified\\\hline
$\tau_{v,s}^{R,A}; \gamma_{R,A}(t); \Gamma_{R,A}(s,f)$ & 
Remaining time to next modification ;
modification rate;
expected number of modification events \\
$\ell_{v}^{R,A}$ & 
Relative exit rate\\
$P_{u,v}^{R,A}(s,f); q_{u,v}^{R,A}$ & 
Transition probability; relative transition rate\\\hline
$\Delta A; \delta; \sigma^{2}$ & 
Change to a value of $A$ in a random-walk update
event; expected value of change; variance of change\\\hline
$C_{R,\text{u}}(s,f);C_{R,\text{o}}(s,f);C_{R}(t);$ & 
Transription cost; obsolescence cost; total cost\\
$\iota_{r,A}(s,f); \iota_{R,A}(s,f); \iota_{r}(s,f)$ & 
Contribution to obsolescence of: $r$ via $A$; $A$; $r$\\
$\hat{\iota}_{R,A}^{\modification}(s,f); 
\hat{\iota}_{R}^{\deletion}(s,f);
\hat{\iota}_{R}^{\medspace\insertion}(s,f)$ & 
Expected obsolescence cost
due to: modification; deletion; insertion\\
$\hat{\iota}_{R,A,u}^{\modification}(s,f)$ & 
Expected obsolescence cost due to modification to the value $u$\\
$c_{u,v}^{R,A}$ & 
Elements of a cost matrix\\\hline
\end{tabular}
}
\caption{List of Symbols.}
\label{tab:listssym}
\end{table}

\renewcommand{\arraystretch}{1}

\section{Modeling insertions and deletions}

\label{sec:estcard}This section introduces the stochastic models
for insertions and deletions. Section
\ref{sec:insertion} discusses insertions, 
while deletions are discussed in section
\ref{sec:deletions}. Section \ref{sec:combinedinsertdelete} 
combines the effect of
insertions and deletions on a relation's cardinality. We conclude 
with a discussion of non-exponential life spans in Section
\ref{sec:nonexplife}. We defer discussing model
validation until Section \ref{sec:verify}.

\subsection{Insertion}
\label{sec:insertion}
We use a nonhomogeneous Poisson process~\cite{ROSS80,TK94}
with instantaneous arrival rate $\lambda_{R}:\Re\rightarrow\lbrack0,\infty)$
to model the occurrence of \emph{insertion events} into $R$. That is, the
number of insertion events occurring in any interval $(s,f]$ is a Poisson
random variable with expected value $\Lambda_{R}(s,f)=\int_{s}^{f}\lambda
_{R}(t)\hspace{0.15em}dt.$ A homogeneous Poisson process may be considered as
the special case where $\lambda_{R}(t)$ is equal to a constant $\lambda_{R}>0$
for all $t$, yielding $\Lambda_{R}(s,f)=\int_{s}^{f}\lambda_{R}(t)\hspace
{0.15em}dt=\int_{s}^{f}\lambda_{R}\hspace{0.15em}dt=\lambda_{R}\cdot(f-s)$.

We now consider the interarrival time distribution of the nonhomogeneous
Poisson process. \ We first define the nonhomogeneous exponential
distribution, as follows:

\begin{definition}
[Nonhomogeneous exponential distribution]\label{def:nhexp}
Let $\phi:\Re\rightarrow\lbrack0,\infty)$ be a integrable
function. Given some $s\in\Re$, a random variable $V$ is said to have a
\emph{nonhomogeneous exponential} distribution (denoted by $V\sim
\expdistrib_{s}(\phi(\cdot))$) if $V$'s density function is
\[
p(\tau)=\left\{
\begin{array}
[c]{ll}%
{\displaystyle\phi(s+\tau)\exp\!{\left(  -\!\!\int_{0}^{\tau}\!\!\!\phi
(s+u)\hspace{0.15em}du\right)  }}, & \tau\geq0\\
0, & \tau<0.
\end{array}
\right.
\]
\end{definition}

It is worth noting that if $\phi(t)$ is constant, $p(\tau)$ is just a standard
exponential distribution. We shall now show that, as with homogeneous Poisson
processes, the interarrival time of insertion events is distributed like an
exponential random variable, $L_{R,s}$, but with a time-varying density function.

\begin{lemma}
\label{lem:interarrival}At any time $s$, the amount of time $L_{R,s}$
to the next insertion event is distributed like $\expdistrib_{s}(\lambda
_{R}(\cdot))$. The probability of an insertion event occurring during $(s,f]$
is $\probop
\!\{L_{R,s}<f-s\}=1-e^{-\Lambda_{R}(s,f)}$.
\end{lemma}

\begin{proof}
\noindent Let $\{N(t),t\geq0\}$ be a nonhomogeneous Poisson process with
intensity function $\lambda_{R}(t)$, which implies $\probop\!\left\{
{N(f)-N(s)=0}\right\}  =e^{-\Lambda_{R}(s,f)}$. Now, the chance that no new
tuple was inserted during $(s,f]$ is the same as the chance that the process
$N(\cdot)$ has no arrivals during $(s,f]$, that is, $e^{-\Lambda_{R}(s,f)}$.
The chance that a new tuple was inserted during $(s,f]$ is just the complement
of the chance of no arrivals, namely,
\[
\probop\!\left\{  L_{R,s}{<f-s}\right\}  =\{N(f)-N(s)\geq
1\}=1-P\{N(f)-N(s)=0\}=1-e^{-\Lambda_{R}(s,f)}.
\]
Taking the derivative of this expression with respect to $f$ and making a
change of variables, the probability density of the time until the next
insertion from time $s$ is $p(\tau)=\lambda_{R}(s+\tau)e^{-\Lambda
_{R}(s,s+\tau)}$. Thus, $L_{R,s}\sim\expdistrib_{s}(\lambda_{R}(\cdot))$.
\end{proof}

At insertion event $i$, a random number of tuples $\Delta_{R,i}^{+}$ are
inserted, allowing us to model bulk insertions. A \emph{bulk insertion} is the
simultaneous arrival of multiple tuples, and may occur because the tuples are
related, or because of limitations in the implementation of the server. For
example, e-mail servers may process an input stream periodically, resulting in
bulk updates of a mailbox. Assuming that the $\{\Delta_{R,i}^{+}\}$ are
independent and identically distributed (IID), then the stochastic process
$\{B_{R}(t),t\geq0\}$ representing the cumulative number of insertions through
time $t$ is a \emph{compound Poisson} process (\emph{e.g.}, \cite{ROSS95}, pp.
87-88). We let $B_{R}(s,f)$ denote the number of insertions falling into the
interval $(s,f]$. The expected number of inserted tuples during $(s,f]$ may be
computed via $\expecop\!\left[  {B}_{R}{(s,f)}\right]  =\int_{s}%
^{f}\!\!\lambda_{R}(t)\expecop
\!\left[  {\Delta}_{R}^{+}\right]  \hspace{0.15em}dt=\expecop\left[  {\Delta
}_{R}^{+}\right]  \int_{s}^{f}\!\!\lambda_{R}(t)\!\hspace{0.15em}%
dt=\expecop\!\left[  {\Delta}_{R}^{+}\right]  \Lambda_{R}(s,f).$ 
Here, $\Delta_{R}^{+}$ represents a generic random variable distributed like
the $\{\Delta_{R,i}^{+}\}$.

We now consider three simple cases of this model:

\paragraph{General nonhomogeneous Poisson process:}
Assume that $\expecop\!\left[  {\Delta_{R}^{+}}\right]  =1$. The expected
number of insertions simplifies to $\expecop\!\left[  {B}_{R}{(s,f)}\right]
=\expecop
\!\left[  {\Delta}_{R}^{+}\right]  \Lambda_{R}(s,f)=1\cdot\Lambda
_{R}(s,f)=\Lambda_{R}(s,f)$.

\paragraph{Homogeneous Poisson process:}
Assume once more that $\expecop\!\left[  {\Delta_{R}^{+}}\right]  =1$. Assume
further that $\lambda_{R}(t)$ is a constant function, that is, $\lambda
_{R}(t)=\lambda_{R}$ for all times $t$. In this case, as shown above,
$\Lambda_{R}(s,f)$ takes on the simple form of $\lambda_{R}\cdot(f-s)$. Thus,
$\expecop\!\left[  {B}_{R}{(s,f)}\right]  =\Lambda_{R}(s,f)=\lambda_{R}%
\cdot(f-s)$. The interarrival times are distributed as $\expdistrib
(\lambda_{R})$, the exponential distribution with parameter $\lambda_{R}$.

\paragraph{Recurrent piecewise-constant Poisson process:}
A simple kind of nonhomogeneous Poisson process can be built out of
homogeneous Poisson processes that repeat in a cyclic pattern. Given some
length of time $T$, such as one day or one week, suppose that the arrival rate
function $\lambda_{R}(t)$ of the recurrent Poisson process repeats every $T$
time units, that is, $\lambda_{R}(t)=\lambda_{R}\!\left(  t-T\!\left\lfloor
{t}/{T}\right\rfloor \right)  $ for all $t$. Furthermore, the interval
$\left[  0,T\right)  $ is partitioned into a finite number of subsets
$J_{1},\ldots,J_{K}$, with $\lambda_{R}(t)$ constant throughout each $J_{k}$,
$k=1,\ldots,K$. Finally, each $J_{k}$ is in turn composed of a finite number
of half-open intervals of the form $[s,f)$. For instance, $T$ might be one
day, with $K=24$ and $J_{1}=[0\text{:}00,1\text{:}00),J_{2}=[1\text{:}%
00,2\text{:}00),\ldots,J_{24}=[23\text{:}00,0\text{:}00)$. As another simple
example, $T$ might be one week, and $K=2$. The subset $J_{1}$ would consist of
a firm's normal hours of operation, say $[9$:$00,18$:$00)$ for each weekday,
and $J_{2}=[0,T)\backslash J_{1}$ would denote all ``off-hour'' times.
Formalisms like those of~\cite{NIEZETTE92} could also be used to describe such
processes in a more structured way. We term this class of Poisson processes to
be \emph{recurrent piecewise-constant} --- abbreviated \emph{RPC}.

It is worth noting that, in client-server environments, the 
insertion model should typically be formed from the client's
point of view. Therefore, if the server keeps a database from which many
clients transcribe data, the modeling of insertions for a given client should
only include the part of the database the client actually transcribes.
Therefore, if a ``road warrior'' is interested only in new orders for the
08904 zip code area, the insertion model for that client should concentrate on
that zip code, ignoring the arrival orders from other areas.

\subsubsection{The complexity of computing $\Lambda_{R}(s,f)$}

\label{sec:lambdacomplex}$\Lambda_{R}(s,f)$, the Poisson
expected value, is computed by integrating the model parameter $\lambda
_{R}(t)$ over the interval $[s,f]$. Standard numerical methods allow rapid
approximation of this definite integral even if no closed formula is known for
the indefinite integral. However, the complexity of this calculation depends
on the information-theoretic properties of $\lambda_{R}(t)$~\cite[Section
1]{TRAUB98}.

For our purposes, however, simple models of $\lambda_{R}(t)$ are likely to
suffice. For example, if $\lambda_{R}(t)$ is a polynomial of degree $d
\geq 0$,
the integration can be performed in $\bigO(d+1)$ time. Consider next a
piecewise-polynomial Poisson process: the time line is divided into intervals
such that, in each time interval, $\lambda_{R}(t)$ can be written as a
polynomial. The complexity of calculating $\Lambda_{R}(s,f)$ in this case is
$\bigO(n(d+1))$, where $n$ is the number of segments in the time interval $(s,f]$,
and $d$ is the highest degree of the $n$ polynomials.

Further suppose that the piecewise-polynomial process is recurrent in a
similar manner to the RPC process, that is, given some fixed time interval
$T$, $\lambda_{R}(t)=\lambda_{R}\!\left(  t-T\!\left\lfloor {t}/{T}%
\right\rfloor \right)  $ for all $t$. Note that the RPC Poisson process is the
special case of this model in which $d=0$. If there are $c$ segments in the
interval $[0,T]$, then the complexity of calculating $\Lambda_{R}(s,f)$
becomes $\bigO(c(d+1))$, regardless of the length of the interval $[s,f]$. This
reduction occurs because, for all intervals of the form $[kT,(k+1)T]\subseteq
[s,f]$ for which $k$ is an integer, the integral $\int_{kT}^{(k+1)T}%
\lambda_{R}(t)\hspace{0.15em} dt$ is equal to $\int_{0}^{T}\lambda
_{R}(t)\hspace{0.15em} dt$, which only needs to be calculated once.

In Section \ref{sec:verify}, we demonstrate the usefulness
of the RPC model for one specific application. We hypothesize that a recurrent
piecewise-polynomial process of modest degree (for example, $d$=3) will be
sufficient to model most systems we are likely to encounter, and so the
complexity of computing $\Lambda_{R}(s,f)$ should be very manageable.

\subsection{Deletion}
\label{sec:deletions}We allow for two distinct deletion mechanisms. First, we
assume individual tuples have their own intrinsic stochastic life spans.
Second, we assume that tuples are deleted to satisfy referential integrity
constraints when tuples in other relations are deleted. 
These two
mechanisms are combined in a tuple's overall probability of being deleted. 
Let $R$ and $S$ be two
relations such that $\mathcal{K}(S)$ is a foreign key of $S$ in $R$. We refer
to $S$ as a \emph{primary relation} of $R$. Consider the directed multigraph
$G$ whose vertices consist of all relations $R$ in the database, and whose
edges are of the form $\langle R,S\rangle$, where $S$ is a primary relation of
$R$. The number of edges $\langle R,S\rangle$ is the number of foreign keys of
$S$ in $R$ for which integrity constraints are enforced. We assume that $G(R)$
has no directed cycles. Let $G(R)$ denote the subgraph of $G$ consisting of
$R$ and all directed paths starting at $R$. We denote the vertices of this
subgraph by $S(R)$.

\begin{figure}
[ptb]
\begin{center}
\epsfig{file=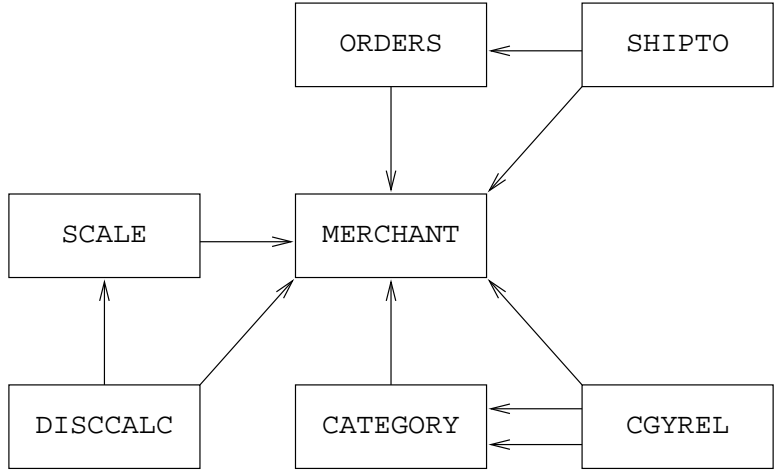}
\caption{A partial multigraph of the case study.}%
\label{fig:multigraph}
\end{center}
\end{figure}

\begin{example}
[Referential integrity constraints in Net.Commerce]
IBM's Net.Commerce is supported by a DB2 database with
about a hundred relations interrelated through foreign keys. For demonstration
purposes, consider a sample of seven relations in the Net.Commerce database.
Figure \ref{fig:multigraph} is a pictorial
representation of the multigraph $G$ of these seven relations. The
\texttt{MERCHANT} relation provides data about merchant profiles, the
\texttt{SCALE} and \texttt{DISCCALC} relations are for computing price
discounts, the \texttt{CATEGORY} and \texttt{CGRYREL} relations assist in
categorizing products, and the \texttt{ORDERS} and \texttt{SHIPTO} relations
contain information about orders. The six relations, \texttt{SCALE, DISCCALC,
CATEGORY,} \texttt{CGRYREL, ORDERS, }and \texttt{SHIPTO} have a foreign key to
the \texttt{MERCHANT} relation, through \texttt{MERCHANT}'s primary key
(\texttt{MERFNBR}). Integrity constraints are enforced between the
\texttt{SCALE} relation and the \texttt{MERCHANT} relation, as long as
\texttt{SCALE.SCLMENBR} (the foreign key to \texttt{MERCHANT.MERFNBR}) does
not have the value \texttt{NULL}. That is, unless a \texttt{NULL} value is
assigned to the \texttt{MERCHANT.MERFNBR} attribute, a deletion of a tuple
in \texttt{MERCHANT} results in a deletion of all tuples in \texttt{SCALE}
such that \texttt{SCALE.SCLMENBR$\,=\,$MERCHANT.MERFNBR}. \texttt{DISCCALC}
has a foreign key to the \texttt{SCALE} relation, through \texttt{SCALE}'s
primary key (\texttt{SCLRFNBR}). There are two attributes of \texttt{CGRYREL}
that serve as foreign keys to the \texttt{CATEGORY} relation, through
\texttt{CATEGORY}'s primary key (\texttt{CGRFNBR}). Finally, \texttt{SHIPTO}
contains shipment information of each product in an order, and therefore it
has a foreign key to \texttt{ORDERS} through its primary key (\texttt{ORFNBR}%
). \hspace*{\fill}$\Box$
\end{example}

With regard to intrinsic deletions within a relation, we assume that each
tuple $r\in R(s)$ has a stochastic remaining life span $L_{R,s}^{\intrinsic}$.
This random variable is identically distributed for each $r\in R(s)$, and is
independent of the remaining life span of any other tuple and of $r$'s age at
time $s$ (see Section \ref{sec:nonexplife} for a
discussion of tuples with a non-memoryless life span). Specifically, we will
assume that the chance of $r\in R(t)$ being deleted in the time interval
$[t,t+\Delta t]$ approaches $\mu_{R}(t)\Delta t$ as $\Delta t\rightarrow0$,
for some function $\mu_{R}:\Re\rightarrow\lbrack0,\infty)$. We define
$M_{R}(s,f)=\int_{s}^{f}\mu_{R}(t)\hspace{0.15em}dt$.

\begin{lemma}
$L_{R,s}^{\intrinsic}\thicksim\expdistrib_{s}(\mu_{R}(\cdot))$. The
probability that a tuple $r\in R(s)$ is deleted by time $f$, given that no
corresponding tuple in $S(R)\backslash\{R\}$ is deleted, is $\probop
\!\{{L_{R,s}^{\intrinsic}<f-s\}=}1-e^{-M_{R}(s,f)}$.
\end{lemma}

\begin{proof}
\noindent Let $r\in R(s)$ be a randomly chosen tuple, and assume that no
corresponding tuple to $r$ in $S(R)\backslash\{R\}$ is deleted. The proof is
identical to that of Lemma \ref{lem:interarrival}, replacing
$\lambda_{R}(t)$ with $\mu_{R}(t)$ and $\Lambda_{R}(s,f)$ by $M_{R}(s,f)$.
\end{proof}

\subsubsection{Deletion and referential integrity}
For any $r\in R(s)$ and any relation $S\in S(R)$, we define $w(r,S)$ to be the
number of tuples in $S$ whose deletion would force deletion of $r$ in order to
maintain referential integrity. This value can be between $0$ and the number
of paths from $R$ to $S$ in $G(R)$. For example, if 
$r\in R=\mathtt{CGRYREL}$ of
Figure \ref{fig:multigraph}, then $0\leq
w(r,\mathtt{CATEGORY})\leq2$ and $0\leq w(r,\mathtt{MERCHANT})\leq3$. For
completeness, we define $w(r,R)=1$. Each tuple in $S$ has an independent
remaining lifetime distributed as $\expdistrib
_{s}(\mu_{S}(\cdot))$, and if any of the $w(r,S)$ tuples corresponding to $r$
is deleted, then $r$ must be immediately deleted, to maintain referential integrity constraints. We use $p_{R}(s,f)$ to
denote the probability that a randomly chosen tuple in $R(s)$ survives until
time $f$.

\begin{lemma}
\label{prop:rawdelete} $p_{R}(s,f)=\expecop_{r\in R(s)}\!\!\left[
\exp\!\left(  -\!\sum_{S\in S(R)}\!w(r,S)M_{S}(s,f)\right)  \right]  $, where
$\expecop
_{r\in R(s)}\!\left[  {\cdot}\right]  $ denotes expectation over random
selection of tuples in $R(s)$.
\end{lemma}

\begin{proof}
\noindent Considering all $S\in S(R)$, and using the well-known fact that if
$L_{i}\sim\expdistrib
_{s}(\mu_{i}(\cdot))$ for $i=1,\ldots,k$ are independent, then
\begin{equation}
\min\!\left\{  L_{1},\ldots,L_{k}\right\}  \sim\expdistrib_{s}\!\left(
\sum_{i=0}^{k}\mu_{i}(\cdot)\right)  ,\label{eq:combineexp}%
\end{equation}
we conclude that the remaining lifetime of $r$ (denoted $L_{R,s}$) has a
nonhomogeneous exponential distribution with intensity function $\sum_{S\in
S(R)}w(r,S)\mu_{S}(\cdot)$. The probability of a given tuple $r\in R(s)$
surviving through time $f$ is thus
\[
\exp\negthickspace\left(  -\int_{s}^{f}\!\!\left(  \sum_{S\in S(R)}%
\!\!\!w(r,S)\mu_{S}(t)\right)  dt\right)  =\exp\negthickspace
\left(  -\!\!\!\sum_{S\in S(R)}\!\!\!w(r,S)M_{S}(s,f)\right)  ,
\]
and the probability that a randomly chosen tuple in $R(s)$ survives until time
$f$ is therefore
\begin{equation}
p_{R}(s,f)=\expecop_{r\in R(s)}\!\!\left[  \exp\!\!\left(  -\!\!\!\sum_{S\in
S(R)}\!\!\!w(r,S)M_{S}(s,f)\right)  \right]  .\label{eq:prdef}%
\end{equation}
\hspace*{\fill}\hspace*{\fill}
\end{proof}

The complexity analysis of integrating $\mu_{S}(t)$ over time to
obtain $M_S(s,f)$
is similar to
that of Section \ref{sec:lambdacomplex}. However, the
computation required by Lemma \ref{prop:rawdelete} may be prohibitive, in the
most general case, because it requires knowing the empirical distribution of
the $w(r,S)$ over all $r\in R(s)$ for all $S\in S(R)$. This empirical
distribution can be computed accurately by computing for each tuple, upon
insertion, the number of tuples in any $S\in S(R)$ with a comparable foreign
key, using either histograms or by directly querying the database. Maintaining
this information requires $\bigO(\left|  {R(s)}\right|  \left|  {S(R)}\right|  )$
space. This complexity can be reduced using a manageably-sized sample from
$R(s)$. Our initial analysis of real-world applications, however, indicates
that in many cases, $w(r,S)$ takes on a much simpler form, in which $w(r,S)$
is identical for all $r\in R(s)$. We term such a typical relationship between
$R$ and $S\in S(R)$ a \emph{fixed multiplicity}, as defined next:

\begin{definition}
The pair $\langle R,S\rangle$, where $S\in S(R)$, has \emph{fixed
multiplicity} if $w(r,S)$ is identical for all tuples in $R$. In this case, we
denote its common value by $w(R,S)$. \hspace*{\fill}$\Box$
\end{definition}

\begin{example}
[Fixed multiplicies in Net.Commerce] Consider the example
multigraph of
Figure \ref{fig:multigraph}. Both \texttt{DISCALC} and
\texttt{SCALE} reference \texttt{MERCHANT}. It is clear that the discount
calculation of a product (as stored in \texttt{DISCALC}) cannot reference a
different merchant than \texttt{SCALE}. The only exception is when the foreign
key in \texttt{SCALE} is assigned with a null value. If this is the case,
however, there is only a single tuple in \texttt{MERCHANT} whose deletion
requires the deletion of a tuple in \texttt{DISCALC}. Thus, for any tuple
$r\in$~\texttt{DISCALC}, $w(r,\mbox{\tt\em SCALE})=w(r,\mbox{\tt\em
MERCHANT})=1$ and therefore $\langle\mbox{\tt\em DISCALC},\mbox{\tt\em
SCALE}\rangle$ and $\langle\mbox{\tt\em DISCALC},\mbox{\tt\em MERCHANT}%
\rangle$ both have fixed multiplicity of $1$. Now consider \texttt{CGRYREL}.
Since each tuple in \texttt{CGRYREL} describes the relationship between a
category and a subcategory, it is clear that its two foreign keys to
\texttt{CATEGORY} must always have distinct values. Thus, $\langle\mbox{\tt\em
CGRYREL},\mbox{\tt\em CATEGORY}\rangle$ has a fixed multiplicity, and
$w(\mbox{\tt\em CGRYREL},\mbox{\tt\em CATEGORY})=2$.\hspace*{\fill}$\Box$
\end{example}

As the following lemma shows, fixed multiplicities permit great simplification
in computing $p_{R}(s,f)$.

\begin{lemma}
\label{prop:fixed} If $\langle R,S\rangle$ has fixed multiplicity for all
$S\in S(R)$, $p_{R}(s,f)=\exp(-\widetilde{M}_{R}(s,f))$, where $\widetilde
{M}_{R}(s,f)=\int_{s}^{f}\tilde{\mu}_{R}(t)\hspace{0.15em}dt$ and $\tilde{\mu
}_{R}(t)=\!\!\sum_{S\in S(R)}w(R,S)\mu_{S}(t)$.
\end{lemma}

\begin{proof}%
\begin{align*}
p_{R}(s,f) &  =\expecop_{r\in R(s)}\!\!\left[  \exp\!\!\left(  -\!\!\!\sum
_{S\in S(R)}\!\!\!w(r,S)M_{S}(s,f)\right)  \right]  \\
&  =\expecop_{r\in R(s)}\!\!\left[  \exp\negthickspace\left(  -\int_{s}%
^{f}\!\!\left(  \sum_{S\in S(R)}\!\!\!w(r,S)\mu_{S}(t)\right)  dt\right)
\right]  \\
&  =\expecop_{r\in R(s)}\!\!\left[  \exp\negthickspace\left(  -\int_{s}%
^{f}\!\!\left(  \sum_{S\in S(R)}\!\!\!w(R,S)\mu_{S}(t)\right)  dt\right)
\right]  \\
&  =\exp\negthickspace\left(  -\!\!\!\sum_{S\in S(R)}\!\!\!\left(
w(R,S)\int_{s}^{f}\!\!\mu_{S}(t)\hspace{0.15em}dt\right)  \right)  \\
&  =\exp\negthickspace{\left( - \!\!\! \sum_{S\in S(R)}%
\!\!\! w(R,S)M_S(s,f)\right)}.
\end{align*}
\end{proof}

Since $w(R,S)$ is fixed and constant over time, no additional statistics need
to be collected for it. As a final note, it is worth noting that in certain
situations, another alternative may also be available. Let $\{N_{R}%
(t),t\geq0\}$ be a nonhomogeneous Poisson process with intensity function
$\hat{\mu}_{R}(t)$, modeling the occurrence of \emph{deletion events} in $R$.
At deletion event $i$, a random number $\Delta_{i}^{-}$ tuples are deleted
from $R$. Generally speaking, this kind of model cannot be accurate, since it
ignores that each deletion causes a reduction in the number of remaining
tuples, and thus presumably a change in the spacing of subsequent deletion
events. However, it may be reasonably accurate for large databases with either
a stable or steadily growing number of tuples, or whenever the time interval
$(s,f]$ is sufficiently small. Statistical analysis of the database log would
be required to say whether the model is applicable. 
If the model is valid, then the
stochastic process $\{D_{R}(t),t\geq0\}$ representing the cumulative number of
deletions through time $t$, can be taken to be a compound Poisson process. The
expected number of deleted tuples during $(s,f]$ may be computed via
\[
\expecop\!\left[  {D_{R}(t)}\right]  =\int_{s}^{f}\!\!\mu_{R}(t)\expecop
\!\left[  {\Delta}^{-}\right]  dt=M_{R}(s,f)\expecop\!\left[  {\Delta}%
^{-}\right]  ,
\]
where $\Delta^{-}$ is a generic random variable distributed like the
$\{\Delta_{i}^{-}\}$.\hspace*{\fill}

\subsection{Tuple survival: the combined effect of insertions and deletions}
\label{sec:combinedinsertdelete} Some tuples inserted during $(s,f]$ may be
deleted by time $f$. Let the random variable $X_{R}(s,f)$ denote the number of
tuples inserted during the interval $(s,f]$ that survive through time $f$.
Consider any tuple inserted into $R$ at time $t\in(s,f]$, and denote its
chance of surviving through time $f$ by $\hat{p}_{R}(t,f)$. For any $S\in
S(R)$ and $t\in(s,f]$, let $W(R,S,t)$ be a random variable denoting the value
of $w(r,S)$, given that $r$ was inserted into $R$ at time $t$. Let $W(R,t)$
denote the random vector, of length $\left|  {S(R)}\right|  $, formed by
concatenating the $W(R,S,t)$ for all $S\in S(R)$.

\begin{lemma}
\label{prop:insertsurvival} $\hat{p}_{R}(t,f)=\expecop_{W(R,t)}\!\left[
{\exp\!\left(  -\sum_{S\in S(R)}W(R,S,t)M_{S}(t,f)\right)  }\right]  $. When
$\langle R,S\rangle$ has fixed multiplicity for all $S\in S(R)$, then $\hat
{p}_{R}(t,f)=p_{R}(t,f)=\exp(-\widetilde{M}_{R}(t,f))$.
\end{lemma}

\begin{proof}
\noindent Let $L_{R,t}$ denote the lifetime of a tuple inserted into $R$ at
time $t$. Similarly to the proof of Lemma \ref{prop:rawdelete}, we know that
$L_{R,t}\thicksim\expdistrib_{t}(\sum_{S\in S(R)}W(R,S,t)\mu_{S}(\cdot))$. The
probability such a tuple survives through time $f$ is the random quantity
\[
\exp\negthickspace\left(  -\int_{t}^{f}\!\!\left(  \sum_{S\in S(R)}%
\!\!\!W(R,S,t)\mu_{S}(\tau)\right)  \hspace{0.15em}d\tau\right)
=\exp\negthickspace
\left(  -\!\!\!\sum_{S\in S(R)}\!\!\!W(R,S,t)M_{S}(t,f)\right)  .
\]
Considering all the possible elements of the vector $W(R,t)$, we then obtain
\[
\hat{p}_{R}(t,f)=\expecop_{W(R,t)}\!\!\left[  \exp\!\!\left(  \!-\!\!\!\!\sum
_{S\in S(R)}\!\!\!W(R,S,t)M_{S}(t,f)\right)  \right]  ,
\]

Assume now that $\langle R,S\rangle$ has fixed multiplicity for all $S\in
S(R)$. Consequently, we replace $W(R,S,t)$ with $w(R,S)$. Drawing on the proof
of the previous lemma,
\begin{align*}
\hat{p}_{R}(t,f) &  =\expecop_{W(R,t)}\!\!\left[  \exp\!\!\left(
\!-\!\!\!\!\sum_{S\in S(R)}\!\!\!w(R,S)M_{S}(t,f)\right)  \right]  \\
&  =\exp\negthickspace\left(  -\int_{t}^{f}\!\!\left(  \sum_{S\in
S(R)}\!\!\!w(R,S)\mu_{S}(\tau)\right)  \hspace{0.15em}d\tau\right)  \\
&  =p_{R}(t,f)\text{.}%
\end{align*}
\hspace*{\fill}
\end{proof}

\noindent The following proposition establishes the formula for the expected
value of ${X_{R}(s,f)}$.

\begin{proposition}
$\expecop\!\left[  {X_{R}(s,f)}\right]  =\widetilde{\Lambda}_{R}%
(s,f)\expecop\!\left[  {\Delta_{R}^{+}}\right]  $, where $\widetilde{\Lambda
}_{R}(s,f)=\int_{s}^{f}\lambda_{R}(t)\hat{p}_{R}(t,f)\hspace{0.15em}dt$. In
the simple case where each insertion involves exactly one tuple,
$X_{R}(s,f)\sim\poissondistrib(\widetilde{\Lambda}_{R}(s,f))$.
\end{proposition}

\begin{proof}
\noindent Let $N$ be the number of insertion events in $(s,f]$, and let their
times be $\{T_{1},T_{2},\ldots,T_{N}\}$. Suppose that $N=n$ and that insertion
event $i$ happens at time $t_{i}\in(s,f]$. Event $i$ inserts a random number
of tuples $\Delta_{R,i}^{+}$, each of which has probability $\hat{p}_{R}%
(t_{i},f)$ of surviving through time $f$. Therefore, the expected number of
tuples surviving through $f$ from insertion event $i$ is $\expecop\!\left[
{\Delta_{R}^{+}}\right]  \hat{p}_{R}(t_{i},f)$. Consequently,
\[
\expecop\!\left[  {X_{R}(s,f)\;\big|\;N=n,T_{1}=t_{1},T_{2}=t_{2},\ldots
,T_{n}=t_{n}}\right]  =\expecop\!\left[  {\Delta_{R}^{+}}\right]  \sum
_{i=1}^{n}\hat{p}_{R}(t_{i},f).
\]
Next, we recall, given that $N=n$, that the times $T_{i}$ of the insertion
events are distributed like $n$ independent random variables with probability
density function $\lambda_{R}(t)/\Lambda_{R}(s,f)$ on the interval $(s,f]$.
Thus,
\begin{align*}
\expecop\!\left[  {X_{R}(s,f)\;\big|\;N=n}\right]   &  =\expecop_{T_{1}%
,\ldots,T_{n}}\!\left[  {\expecop\!\left[  {\Delta_{R}^{+}}\right]  \sum
_{i=1}^{n}\hat{p}_{R}(T_{i},f)}\right]  \\
&  =\expecop\!\left[  {\Delta_{R}^{+}}\right]  \sum_{i=1}^{n}\left(  \int
_{s}^{f}\!\!\hat{p}_{R}(t,f)\frac{\lambda_{R}(t)}{\Lambda_{R}(s,f)}%
\hspace{0.15em}dt\right)  \\
&  =n\left(  \frac{\expecop\!\left[  {\Delta_{R}^{+}}\right]  \widetilde
{\Lambda}_{R}(s,f)}{\Lambda_{R}(s,f)}\right)  .
\end{align*}
Finally, removing the conditioning on $N=n$, we obtain
\begin{align*}
\expecop\!\left[  {X_{R}(s,f)}\right]   &  =\expecop_{N}\!\left[  {N\left(
\frac{\expecop\!\left[  {\Delta_{R}^{+}}\right]  \widetilde{\Lambda}_{R}%
(s,f)}{\Lambda_{R}(s,f)}\right)  }\right]  \\
&  =\Lambda_{R}(s,f)\left(  \frac{\expecop\!\left[  {\Delta_{R}^{+}}\right]
\widetilde{\Lambda}_{R}(s,f)}{\Lambda_{R}(s,f)}\right)  \\
&  =\expecop\!\left[  {\Delta_{R}^{+}}\right]  \widetilde{\Lambda}_{R}(s,f).
\end{align*}

In the case that $\Delta_{R}^{+}$ is always $1$, we may use the notion of a
\emph{filtered} Poisson process: if we consider only tuples that manage to
survive until time $f$, the chance of a single insertion in time interval
$[t,t+\Delta t]$ no longer has the limiting value $\lambda_{R}(t)\Delta t$,
but instead $\lambda_{R}(t)\hat{p}_{R}(t,f)\Delta t$. Therefore, the insertion
of surviving tuples can be viewed as a nonhomogeneous Poisson process with
intensity function $\lambda_{R}(t)\hat{p}_{R}(t,f)$ over the time interval
$(s,f]$, so $X_{R}(s,f)\sim\poissondistrib(\widetilde{\Lambda}_{R}(s,f))$.
\end{proof}

In the general case, the computation of $\widetilde{\Lambda}_{R}(s,f)$ will
require approximation by numerical integration techniques; the complexity of
this calculation will depend on the information-theoretic properties of
$\lambda_{R}(\cdot)$ and the $\mu_{S}(\cdot)$, $S\in S(R)$, but is unlikely to
be burdensome if these functions are reasonably smoothly-varying. In one
important special case, however, the complexity of computing $\widetilde
{\Lambda}_{R}(s,f)$ is essentially the same as that of calculating
$\Lambda_{R}(s,f)$: suppose that for some constants $\alpha(R,S)$, $S\in
S(R)$, one has that $\mu_{S}(t)=\alpha(R,S)\lambda_{R}(t)$ for all $t$. That
is, the general insertion and deletion activity level of the relations in
$S(R)$ all vary proportionally to some common fluctuation pattern. In this
case, we have $\widetilde{\mu}_{R}(t) = \alpha(R)\lambda_{R}(t)$ and
$\widetilde{M}_{R}(s,f) = \alpha(R)\Lambda_{R}(s,f)$ for all $t,s,f$, where
$\alpha(R) = \sum_{S\in S(R)} \alpha(R,S)$. Making a substitution
$u(t)=\Lambda_{R}(t,f)$, we have:
\begin{align*}
\widetilde{\Lambda}_{R}(s,f)  &  = \int_{s}^{f} \!\! \lambda_{R}(t)
\exp(-\alpha(R)\Lambda_{R}(t,f)) \hspace{0.15em} dt\\
&  = \int_{s}^{f} \! \left(  \frac{-d\Lambda_{R}(t,f)}{dt} \right)
\exp(-\alpha(R)\Lambda_{R}(t,f)) \hspace{0.15em} dt\\
&  = \int_{s}^{f} \!\! - \exp(-\alpha(R)u(t)) \hspace{0.15em} du(t)\\
&  = - \int_{u(s)}^{u(f)} \!\! e^{-\alpha(R)u} \hspace{0.15em} du\\
&  = \frac{1}{\alpha(R)} \left(  1 - e^{-\alpha(R)\Lambda(s,f)} \right)  ,
\end{align*}
so $\widetilde{\Lambda}_{R}(s,f)$ can be calculated directly from
$\Lambda(s,f)$.

We define the random variable $Y_{R}(s,f)$ to be the number of tuples in
$R(s)$ that survive through time $f$.

\begin{proposition}
$\expecop\!\left[  {Y_{R}(s,f)}\right]  =p_{R}(s,f)\left|  {R(s)}\right|  $
and $\expecop\!\left[  {\left|  {R(f)}\right|  }\right]  =p_{R}(s,f)\left|
{R(s)}\right|  +\widetilde{\Lambda}_{R}(s,f)\expecop\!\left[  {\Delta}_{R}%
^{+}\right]  $.
\end{proposition}

\begin{proof}
\noindent Each tuple in $R(s)$ has a survival probability of $p_{R}(s,f)$,
which yields that $\expecop
\!\left[  {Y_{R}(s,f)}\right]  =p_{R}(s,f)\left|  {R(s)}\right|  $. By the
definitions of $Y_{R}(s,f)$ and $X_{R}(s,f)$, one has that
\[
\left|  {R(f)}\right|  =Y_{R}(s,f)+X_{R}(s,f),
\]
so therefore
\[
\expecop\!\left[  {\left|  {R(f)}\right|  }\right]  =\expecop\!\left[
{Y_{R}(s,f)}\right]  +\expecop\!\left[  {X_{R}(s,f)}\right]  =p_{R}%
(s,f)\left|  {R(s)}\right|  +\widetilde{\Lambda}_{R}(s,f)\expecop\!\left[
{\Delta_{R}}^{+}\right]  .
\]
\hspace*{\fill}\hspace*{\fill}\hspace*{\fill}
\end{proof}

In cases where deletions may also be accurately modeled as a compound Poisson
process, we have
\begin{align*}
\expecop\!\left[  {\left|  {R(f)}\right|  }\right]   &  =\expecop\!\left[
{\left|  {R(s)}\right|  }\right]  +B_{R}(s,f)-D_{R}(s,f)\\
&  =\left|  {R(s)}\right|  +\Lambda_{R}(s,f)\expecop\!\left[  {\Delta}%
^{+}\right]  -M_{R}(s,f)\expecop\!\left[  {\Delta}^{-}\right]  .
\end{align*}

\begin{example}
[The homogeneous case]Assume that $\expecop\!\left[  {\Delta_{R}^{+}}\right]
=1$, that $\langle R,S\rangle$ has fixed multiplicity for all $S\in S(R)$, and
furthermore $\lambda_{R}(t)$ and $\mu_{S}(t)$, for all $S\in S(R)$, are
constant functions, that is, $\lambda_{R}(t)=\lambda_{R}$ for all times $t$
and $\mu_{S}(t)=\mu_{S}$ for all $S\in S(R)$ and times $t$. Then $\Lambda
_{R}(s,f)=\lambda_{R}\cdot(f-s)$ and $M_{R}(s,f)=\mu_{R}\cdot(f-s)$. Thus,
letting $\tilde{\mu}_{R}=\sum_{S\in S(R)}w(R,s)\mu_{S}$,
\[
\widetilde{\Lambda}_{R}(s,f)=\int_{s}^{f}\!\!\lambda_{R}e^{-\tilde{\mu}%
_{R}(t-s)}\hspace{0.15em}dt=\frac{\lambda_{R}}{\tilde{\mu}_{R}}\left(
1-e^{-\tilde{\mu}_{R}(f-s)}\right)  ,
\]
and assuming ${\Delta}_{R,i}^{+}=1$ for all $i>0$,
\[
\expecop\!\left[  {\left|  {R(f)}\right|  }\right]  =\left|  {R(s)}\right|
e^{-\tilde{\mu}_{R}(f-s)}+\frac{\lambda_{R}}{\tilde{\mu}_{R}}\left(
1-e^{-\tilde{\mu}_{R}(f-s)}\right)  =\frac{\lambda_{R}}{\tilde{\mu}_{R}%
}+e^{-\tilde{\mu}_{R}(f-s)}\left(  \left|  {R(s)}\right|  -\frac{\lambda_{R}%
}{\tilde{\mu}_{R}}\right)  .
\]
\hspace*{\fill}$\Box$
\end{example}

\subsection{Tuples with non-exponential life spans}
\label{sec:nonexplife}We now consider the possibility
that tuples in $R$ have a stochastic life span $L_{R}^{\intrinsic}$ that is
not memoryless, but rather has some general cumulative distribution function
$G_{R}$. For example, if tuples in $R$ correspond to pieces of work in process
on a production floor, the likelihood of deletion might rise the longer the
tuple has been in existence. Let us consider a single relation, and thus no
referential integrity constraints. For any tuple $r$, let $b(r)$ denote the
time it was created. We next establish the expected cardinality of $R$ at time
$f$.

\begin{proposition}
In the case that tuples in $R$ have lifetimes with a general cumulative
distribution function $G_{R}$,
\begin{equation}
\expecop\!\left[  {\left|  {R(f)}\right|  }\right]  =\left|  {R(s)}\right|
\expecop_{r\in R(s)}\!\left[  {\frac{1-G_{R}(f-b(r))}{1-G_{R}(s-b(r))}%
}\right]  +\expecop\!\left[  {\Delta}_{R}^{+}\right]  \int_{s}^{f}%
\!\!\lambda_{R}(t)\left(  1-G_{R}(f-t)\right)  \hspace{0.15em}%
dt.\label{eq:nonexperf}%
\end{equation}
\end{proposition}

\begin{proof}
\noindent Let $L_{R}^{\intrinsic}$ denote a generic random variable with
cumulative distribution $G_{R}$. The probability of $r\in R(s)$ surviving
throughout $(s,f]$ is then
\[
\probop\!\left\{  L_{R}^{\intrinsic}\geq f-b(r)\;\big|\;L_{R}^{\intrinsic
}\geq s-b(r)\right\}  =\frac{1-G_{R}(f-b(r))}{1-G_{R}(s-b(r))},
\]
and therefore the expected number of tuples in $R(s)$ that survive through
time $f$ is
\[
\expecop\!\left[  {Y_{R}(s,f)}\right]  =\!\!\!\sum_{r\in R(s)}\!\!\!\left(
\frac{1-G_{R}(f-b(r))}{1-G_{R}(s-b(r))}\right)  =\left|  {R(s)}\right|
\expecop_{r\in R(s)}\!\left[  {\frac{1-G_{R}(f-b(r))}{1-G_{R}(s-b(r))}%
}\right]  .
\]
We now consider a tuple $r$ inserted at some time $t\in(s,f]$. The probability
that such a tuple survives through time $f$ is simply $\hat{p}_{R}%
(t,f)=1-G_{R}(f-t)$. By reasoning similar to the proof of Lemma
~\ref{prop:insertsurvival},
\[
\expecop\!\left[  {X_{R}(s,f]}\right]  =\expecop
\!\left[  {\Delta_{R}}^{+}\right]  \widetilde{\Lambda}_{R}(s,f)=\expecop
\!\left[  {\Delta_{R}}^{+}\right]  \int_{s}^{f\!\!}\lambda_{R}(t)\left(
1-G_{R}(f-t)\right)  \hspace{0.15em}dt.
\]
The conclusion then follows from $\expecop\!\left[  {\left|  {R(f)}\right|
}\right]  =\expecop\!\left[  {Y_{R}(s,f)}\right]  +\expecop\!\left[
{X_{R}(s,f)}]\right.  $
\end{proof}

It is worth noting that, as opposed to the memoryless case presented above,
the calculation of $\expecop\!\left[  {Y_{R}(s,f)}\right]  $ requires
remembering the commit times $b(r)$ of all tuples $r\in R(s)$, or equivalently
the ages of all such tuples. Of course, for large relations $R$, a reasonable
approximation could be obtained by using a manageably-sized sample to
estimate
\[
\expecop_{r\in R(s)}\!\left[  \frac{1-G_{R}(f-b(r))}{1-G_{R}(s-b(r))}
\right]  .
\]
It is likely that the integral in (\ref{eq:nonexperf}) will require general
numerical integration, depending on the exact form of $G_{R}$.

\subsection{Summary}

In this section, we have provided a model for the insertion and deletion of
tuples in a relational database. The immediate benefit of this model is the
computation of the expected relation cardinality ($\expecop\!\left[  {\left|
{R(f)}\right|  }\right]  $), given an initial cardinality and insertion and
tuple life span parameters. Relation cardinality has proven to be an important
property in many database tools, including query optimization and database
tuning. Reasonable assumptions regarding constant multiplicity allow, once
appropriate statistics have been gathered, a rapid computation of
cardinalities in this framework. Section \ref{sec:verify}
elaborates on statistics gathering and model validation.

A note regarding tuples with non-exponential life spans is now warranted. For
the case of a single relation, non-exponential life spans add only a moderate
amount of complexity to our model, namely the requirement to store at least an
approximation of the distribution of tuples ages in $R(s)$. For multiple
relations with referential integrity constraints, 
however, the complexity of dealing with
general tuple life spans is much greater. First, to estimate the cardinality
of $R(f)$, we must keep (approximate) tuple age distributions for all
relations in $S(R)$. Second, because the tuple life span distributions of some
of the members of $S(R)$ are not memoryless, we cannot combine them with a
simple relation like (\ref{eq:combineexp}). Furthermore, in attempting to find
the distribution of the remaining life span of a particular tuple $r\in R(s)$,
it may become necessary to consider the issue of the correlation of ages of
tuples in $R(s)$ with the ages of the corresponding tuples in other relations
of $S(R)$. Because of these complications, we defer further consideration of
non-exponential tuple life spans to future research.

\section{Modeling data modification}
\label{sec:modif}This section describes various ways to model
the modification of the contents of tuples. We start with a general approach,
using Markov chains, followed by several special cases where the amount of
computation can be greatly reduced.

\subsection{Content-dependent updates}

\label{sec:condepup}In this section, we model the modification of the contents of tuples as a
finite-state continuous-time Markov chain, thus assuming dependence on tuples'
previous contents. For each relation $R$, we allow for some (possibly empty)
subset $\mathcal{C}(R)\subset\mathcal{A}(R)$ of its attributes to be subject
to change over the lifetime of a tuple. We do not permit primary key fields to
be modified, that is, $\mathcal{C}(R)\cap\mathcal{K}(R)=\emptyset$.

Attribute values may change at time instants called \emph{transition events},
which are the transition times of the Markov chain. We assume that the spacing
of transition events is memoryless with respect to the age of a tuple
(although it may depend on the time and the current value of the attribute, as
demonstrated below). For any attribute $A$, tuple $r$, time $s$, and value
$v\in\dom A$ with $r.A(s)=v$, the time remaining until the next transition
event for $r.A$ is a random variable $\tau_{v,s}^{R,A}$ with the distribution
$\expdistrib
_{s}(\ell_{v}^{R,A}\gamma_{R,A}(\cdot))$, where $\gamma_{R,A}:\Re
\rightarrow\lbrack0,\infty)$ is a function giving the general instantaneous
rate of change for the attribute, and $\ell_{v}^{R,A}$ is a nonnegative scalar
which we call the \emph{relative exit rate} of $v$. We define $\Gamma
_{R,A}(s,f)=\int_{s}^{f}\gamma_{R,A}(t)\hspace{0.15em}dt$. When a transition
event occurs from state $u\in\dom A$, attribute $A$ changes to $v\in\dom A$
with probability $P_{u,v}^{R,A}$.

Suppose $\mathcal{A}=\{A_{1},A_{2},...,A_{k}\}\subseteq\mathcal{R}$ is an
independently varying set of attributes, and $v=\langle v_{1},v_{2}%
,...,v_{k}\rangle\in\dom\mathcal{A}$ is a compound value. Then the time until
the next transition event for $r.\mathcal{A}$ is $\tau_{v,s}^{R,\mathcal{A}}=\min
\{\tau_{v_{1},s}^{R,A_{1}},\ldots,\tau_{v_{k},s}^{R,A_{k}}\}$. As a rule, we
will assume that the modification processes for the attributes of a relation
are independent, so $\tau_{v,s}^{R,\mathcal{A}}\sim\expdistrib
_{s}(\sum_{i=1}^{k}\ell_{v_{i}}^{R,A_{i}}\gamma_{A_{i},R}(\cdot))$. When the
functions $\gamma_{R,A_{i}}$ are identical for $i=1,\ldots,k$, we define
$\gamma_{R,\mathcal{A}}=\gamma_{R,A_{i}}$ and $\ell_{v}^{R,\mathcal{A}}=\sum_{i=1}^{k}\ell_{v_{i}%
}^{R,A_{i}}$, so $\tau_{v,s}^{R,\mathcal{A}}\sim\expdistrib_{s}(\ell_{v}^{R,\mathcal{A}}%
\gamma_{R,\mathcal{A}}(\cdot))$. To justify the assumption of independence, we note that
coordinated modifications among attributes can be modeled by replacing the
coordinated attributes with a single compound attribute (this technique
requires that the attributes have identical $\gamma_{R,\mathcal{A}}(\cdot)$ functions,
which is reasonable if they change in a coordinated way).

Under these assumptions,
let $\overline{\mathcal{C}}(R)$ denote a partition of $\mathcal{C}(R)$ into
subsets $\mathcal{A}$ such that any two attributes $A_{1},A_{2}\in
\mathcal{C}(R)$ vary dependently iff they are in the same $\mathcal{A}%
\in\overline{\mathcal{C}}(R)$.

\begin{example}
[First alteration time]\label{ex:fat} For a relation $R$ and time
$s$, we define $\Upsilon_{R,s}$ to be the amount of time until the next change
in $R$, be it a tuple insertion, a tuple deletion, or an attribute
modification. Also, for any $S\in S(R)$, let $D(R,S,s)$ denote the number of
tuples in $S(s)$ whose deletion would force the deletion of some tuple in
$R(s)$ $($and therefore $D(R,R,s)=\left|  {R(s)}\right|  )$. The following
proposition establishes the distribution of $\Upsilon_{R,s}$.
\end{example}

\begin{proposition}
$\Upsilon_{R,s}\sim\expdistrib_{s}(\zeta_{R}(\cdot))$, where
\[
\zeta_{R}(t)=\lambda_{R}(t)\;\;+\sum_{S\in S(R)}\!\!\!D(R,S,s)\mu
_{S}(t)\;\;+\sum_{\mathcal{A}\in\overline{\mathcal{C}}(R)}\!h(R,A,s)\gamma
_{R,\mathcal{A}}(t)
\]
and $h(R,A,s)=\sum_{v\in\dom\mathcal{A}}\hat{R}_{\mathcal{A},v}(s)\ell
_{v}^{R,A}$. The probability of any alteration to $R$ in the time interval
$(s,f]$ is $1-e^{-Z_{R}(s,f)}$, where
\[
Z_{R}(s,f)=\Lambda_{R}(s,f)\;\;+\sum_{S\in S(R)}\!\!\!D(R,S,s)M_{S}%
(t)\;\;+\sum_{\mathcal{A}\in\overline{\mathcal{C}}(R)}\!\!\!h(R,A,s)\Gamma
_{R,\mathcal{A}}(s,f)
\]
\end{proposition}

\begin{proof}
Let $\Upsilon_{R,s}^{\insertion}$, $\Upsilon_{R,s}^{\modification}$, and
$\Upsilon_{R,s}^{\deletion}$ be the times until the next insertion,
modification, and deletion in $R$, respectively. From Section
\ref{sec:estcard}, we have that $\Upsilon_{R,s}^{\insertion}\sim\expdistrib
_{s}(\lambda_{R}(\cdot))$. Now, for each $S\in S(R)$, there are $D(R,S,s)$
tuples whose deletion would cause a deletion in $R$. The time until deletion
of any such $r\in S\in S(R)$ is distributed like $\expdistrib_{s}(\mu
_{S}(\cdot))$. The deletion processes for all these tuples are independent
across all of $S(R)$, so we can use (\ref{eq:combineexp}) to conclude that
\[
\Upsilon_{R,s}^{\deletion}\sim\expdistrib_{s}\left(  \sum_{S\in S(R)}%
\!\!\!D(R,S,s)\mu_{S}(\cdot)\right)  .
\]
From the preceding discussion, we have
\[
\Upsilon_{R,s}^{\modification}=\tau_{v,s}^{\mathcal{C}(R),R}\sim\expdistrib
_{s}\left(  \sum_{\mathcal{A}\in\overline{\mathcal{C}}(R)}\ell_{r.\mathcal{A}%
(s)}^{R,\mathcal{A}}\gamma_{R,\mathcal{A}}(\cdot)\right)  .
\]
Since $\Upsilon_{R,s}=\min\{\Upsilon_{R,s}^{\insertion
},\Upsilon_{R,s}^{\modification},\Upsilon_{R,s}^{\deletion}\}$, we therefore
have, again using independence and (\ref{eq:combineexp}), that $\Upsilon
_{R,s}\sim\expdistrib_{s}(\zeta_{R}(\cdot))$, where
\begin{align*}
\zeta_{R}(t) &  =\lambda_{R}(t)\;\;+\sum_{S\in S(R)}\!\!\!D(R,S,s)\mu
_{S}(t)\;\;+\sum_{r\in R(s)}\!\!\left[  \sum_{\mathcal{A}\in\overline
{\mathcal{C}}(R)}\!\!\!\ell_{r.\mathcal{A}(s)}^{R,\mathcal{A}}\gamma
_{R,\mathcal{A}}(t)\right]  \\
&  =\lambda_{R}(t)\;\;+\sum_{S\in S(R)}\!\!\!D(R,S,s)\mu_{S}(t)\;\;+\sum
_{\mathcal{A}\in\overline{\mathcal{C}}(R)}\!\!\!h(R,A,s)\gamma_{R,\mathcal{A}%
}(t).
\end{align*}
Integrating over $(s,f]$ results in
\begin{align*}
Z_{R}(s,f) &  =\int_{s}^{f}\!\zeta_{R}(t)\hspace{0.15em}dt\\
&  =\Lambda_{R}(s,f)\;\;+\sum_{S\in S(R)}\!\!\!D(R,S,s)M_{S}(t)\;\;+\sum
_{\mathcal{A}\in\overline{\mathcal{C}}(R)}\!\!\!h(R,A,s)\gamma_{R,\mathcal{A}%
}(s,f).
\end{align*}
Therefore, the probability of any alteration to $R$ in the time interval
$(s,f]$ is%
\[
\probop\left\{  \Upsilon_{R,s}<f-s\right\}  =1-e^{-Z_{R}(s,f)}%
\]
\hspace*{\fill}
\end{proof}

\noindent\textbf{Example \ref{ex:fat} continued (First alteration
transcription policy)~~} \textit{Suppose the
user wishes to refresh her replica of relation $R$ whenever the
probability that it
contains any inaccuracy exceeds some threshold $\pi$, a tactic we call
the \emph{first
alteration policy}. Then, a refresh is required at time $f$ if $1-e^{-Z_{R}
(s,f)}>\pi$.}\hspace*{\fill}$\Box$

~

Given $\mathcal{A}$, we describe the transition process for the value
$r.\mathcal{A}$ over time by probabilities
\[
P_{u,v}^{R,\mathcal{A}}(s,f)=\probop\!\left\{  r.\mathcal{A}(f)\!=\!v\;\big
|\;r.\mathcal{A}(s)\!=\!u\right\}  ,
\]
for any two values $u,v\in\dom\mathcal{A}$ and times $s<f$. Under the
assumption of independence,
\begin{equation}
P_{u,v}^{R,\mathcal{A}}(s,f)=\prod_{i=1}^{k}P_{u_{i},v_{i}}^{R,A_{i}}(s,f).
\label{eq:probproduct}%
\end{equation}
Given any simple attribute $A$, we define $q_{u,v}^{R,A}$, the \emph{relative
transition rate} from $u$ to $v$, by
\[
q_{u,v}^{R,A}=\ell_{u}^{R,A}P_{u,v}^{R,A}.
\]
Given a set of attributes $\mathcal{A}$ with identical $\gamma_{R,A}(\cdot)$
functions, the compound transition rate $q_{u,v}^{R,A}$ may be computed via
\begin{equation}
q_{u,v}^{R,\mathcal{A}}=\ell_{u}^{R,\mathcal{A}}P_{u,v}^{R,\mathcal{A}%
}=\left(  \sum_{i=1}^{k}\ell_{u_{i}}^{R,A_{i}}\right)  \left(  \prod_{i=1}%
^{k}P_{u_{i},v_{i}}^{R,A_{i}}\right)  . \label{eq:jointexit}%
\end{equation}
Let $Q^{R,A}$ be the matrix of $q_{u,v}^{R,A}$, where $q_{u,u}^{R,A}=-\ell
_{u}^{R,A}$.

\begin{proposition}
The matrix $P^{R,A}(s,f)$ of elements $P_{u,v}^{R,A}(s,f)$ is given by the
matrix exponential formula
\begin{equation}
P^{R,A}(s,f)=\exp\!\left(  \Gamma_{R,A}(s,f)\,Q^{R,A}\right)  =\sum
_{n=0}^{\infty}\frac{\Gamma_{R,A}(s,f)^{n}}{n!}{\left(  Q^{R,A}\right)  }%
^{n}.\label{eq:matrixexp}%
\end{equation}
\end{proposition}

\begin{proof}
\noindent Consider a continuous-time Markov chain on the same state space
$\dom
A$, and with the same instantaneous transition probabilities $P_{u,v}^{R,A}$,
where $u,v\in\dom A$. However, in the new chain, the holding time in each
state $v$ is simply a homogeneous exponential random variable with arrival
rate $\ell_{v}^{R,A}$. We call this system the \emph{linear-time} chain, to
distinguish it from the original chain. Define $\overline{P}_{u,v}^{A,R}(t)$
to be the chance that the linear-time chain is in state $v$ at time $t$, given
that it is in state $u$ at time $0$. Standard results for finite-state
continuous time Markov chains imply that
\[
\overline{P}_{u,v}^{R,A}(t)=\exp\!\left(  t\,Q^{R,A}\right)  =\sum
_{n=0}^{\infty}\frac{t^{n}}{n!}{\left(  Q^{R,A}\right)  }^{n}.
\]
By a transformation of the time variable, we then assert that
\[
P_{u,v}^{R,A}(s,f)=\overline{P}_{u,v}^{R,A}(\Gamma_{R,A}(s,f)),
\]
from which the result follows.
\end{proof}

\begin{example}[Query optimization, revisited]
\label{ex:qopt2}As the following proposition shows, our model
can be used to estimate the histogram of a relation $R$ at time $f$. A query
optimizer running at time $f$ could use expected histograms, calculated in
this manner, instead of the old histograms $\hat{R}_{A}(s)$.
\end{example}

\begin{proposition}
Assume that $w(r,S)$, for all $S\in S(R)$, is independent of the attribute
values $r.A(s)$ for all $A\in\mathcal{C}(R)$. Let $\hat{\omega}_{u}^{R,A}(t)$
denote the probability that $r.A(t)=u$, given that $r$ is inserted into $R$ at
time $t$. Then, for all $v\in\dom A$,
\begin{align}
\expecop\!\left[  {\hat{R}_{A,v}(f)}\right]   &  =p_{R}(s,f)\!\!\!\!\sum
_{u\in\dom A}\!\!\!\!\hat{R}_{A,u}(s)P_{u,v}^{R,A}(s,f)\nonumber\\
&  \quad\quad+\quad\expecop\!\left[  {\Delta_{R}^{+}}\right]  \!\!\!\!\sum
_{u\in\dom A}\!\!\!\left(  \int_{s}^{f}\!\!\hat{\omega}_{u}^{R,A}(t)\hat
{p}_{R}(s,f)\lambda_{R}(t)P_{u,v}^{R,A}(t,f)\intdspace dt
\right)  .\label{eq:query}%
\end{align}
\end{proposition}

\begin{proof}
We first compute the expected number of surviving tuples $r$ whose values
$r.A$ migrate to $v$. Given a value $u\in\dom A$, there are $\hat{R}_{A,u}(s)$
tuples at time $s$ such that $r.A(s)=u$. Using the previous results, the
expected number of these tuples surviving through time $f$ is $\hat{R}%
_{A,u}(s)p_{R}(s,f)$, and the probability of each surviving tuple $r$ having
$r.A(f)=v$ is $P_{u,v}^{R,A}(s,f)$. Using the independence assumption and
summing over all $u\in\dom
A$, one has that the expected numbers of tuples in $R(s)$ that survive through
$f$ and have $r.A(f)=v$ is
\[
\sum_{u\in\dom A}\!\!\!\!\hat{R}_{A,u}(s)p_{R}(s,f)P_{u,v}^{R,A}%
(s,f)=p_{R}(s,f)\!\!\!\!\sum_{u\in\dom A}\!\!\!\!\hat{R}_{A,u}(s)P_{u,v}%
^{R,A}(s,f).
\]
We next consider newly inserted tuples. Recall that $\hat{\omega}_{u}%
^{R,A}(t)$ denotes the probability that $r.A(t)=u$, given that $r$ is inserted
into $R$ at time $t$. Suppose that an insertion occurs at time $t\in(s,f]$.
The expected number of tuples $r$ created at this insertion that both survive
until $f$ and have $r.A(f)=v$ is
\[
\hat{p}_{R}(t,f)\!\!\!\sum_{u\in\dom A}\!\!\!\!{\omega}_{u}^{R,A}%
(t)P_{u,v}^{R,A}(t,f).
\]
By logic similar to Proposition~\ref{prop:insertsurvival}, one may then
conclude that the expected number of newly-inserted tuples that survive
through time $f$ and have $r.A(f)=v$ is
\[
\expecop\!\left[  {\Delta_{R}^{+}}\right]  \!\!\!\!\sum_{u\in\dom
A}\!\!\!\left(  \int_{s}^{f}\!\!\hat{\omega}_{u}^{R,A}(t)\hat{p}%
_{R}(s,f)\lambda_{R}(t)P_{u,v}^{R,A}(t,f)\hspace{0.15em}dt\right)  .
\]
The result follows by adding the last two expressions.
\end{proof}

\noindent\textbf{Example \ref{ex:qopt2} continued~} \emph{We next
consider whether the complexity of calculating (\ref{eq:query}) is
preferable to recomputing the histogram vector ${\hat{R}_{A}(f)}$.
This topic is quite involved and depends heavily on the specific structure of
the database (\emph{e.g.}, the availability of indices) and the
specific application (\emph{e.g.}, the concentration of values in a
small subset of an attribute's domain). In what follows, we lay out
some qualitative considerations in deciding whether calculating
(\ref{eq:query}) would be more efficient than recalculating
${\hat{R}_{A}(f)}$ ``from scratch.''  Experimentation with
real-world application is left for further research.}

\emph{Generally speaking, direct computation of the histogram of an
attribute $A$ (in the absence of an index for $A$) can be done by
either scanning all tuples (although sampling may also be used) or
scanning a modification log to capture changes to the prior histogram
vector ${\hat{R}_{A}(s)}$ during $(s,f]$. Therefore, the 
recomputation can be performed in $\bigO(\min\{\card{R(f)},T(s,f)\})$
time, where $T(s,f)$ denotes the total number of updates during
$(s,f]$. Whenever $\card{R(f)}$ and $T(s,f)$ are both large ---
\emph{i.e.}, the database is large and the transaction load is high
--- the straightforward techniques will be relatively
unattractive. As for the estimation technique, it will probably work
best when $\card{domA}$ is small (for example, for a binary attribute)
or whenever the subset of actually utilized values in the domain is
small. In addition, commercial databases recompute
the entire histogram as a single, atomic task. Formula
(\ref{eq:query}), on the other hand, can be performed on a subset of the
attribute values.  For example, in the case of exact matching (say, a
condition of the form $\mathtt{WHERE}\;A=v$), it is sufficient to
compute $\hat{R}_{A,v}(f)$, rather than the full
${\hat{R}_{A}(f)}$ vector. Finally, it is worth noting that the
computing the expected value 
of ${\hat{R}_{A,v}(f)}$ via (\ref{eq:query}) does not require locking $R$,
while a full histogram recomputation may involve extended periods of
locking.}\hspace*{\fill}$\Box$

~

We next consider the number of tuples in $R(s)$ that have survived through
time $f$ without being modified, which we denote $Y_{R}^{-}(s,f)$. The
expectation of this random variable is
\[
\expecop\!\left[  {Y_{R}^{-}(s,f)}\right]  =p_{R}(s,f)\!\!\!\!
\sum_{v\in\dom{\mathcal{A}}}\!\!\!\!
\hat{R}_{\mathcal{A},v}(s)P_{v,v}^{R,\mathcal{A}}(s,f)
\]
We let $Y_{R}^{+}(s,f)=Y_{R}(s,f)-Y_{R}^{-}(s,f)$ denote the number of tuples
in $R(s)$ that have survived through time $f$ and were modified; it follows
from the linearity of the $\expecop\!\left[  {\cdot}\right]  $ operator that
\[
\expecop\!\left[  {Y_{R}^{+}(s,f)}\right]  =\expecop\!\left[  {Y_{R}%
(s,f)}\right]  -\expecop\!\left[  {Y_{R}^{-}(s,f)}\right]  .
\]

\subsubsection{Complexity analysis of content-dependent updates}

In practice, as with computing a scalar exponential, only a limited number of
terms will be needed to compute the sum (\ref{eq:matrixexp}) to machine
precision. It is worth noting that efficient means of calculating
(\ref{eq:matrixexp}) are a major topic in the field of computational probability.

In the case of a compound attribute $\mathcal{A}=\{A_{1},A_{2},...,A_{k}\}$
with independently varying components, it will be computationally more
efficient to first calculate the individual transition probability matrices
$P^{A_{i},R}(s,f)$ via (\ref{eq:matrixexp}), and then calculate the joint
probability matrix $P_{u,v}^{R,\mathcal{A}}(s,f)$ using (\ref{eq:probproduct}%
), rather than first finding the joint exit rate matrix $Q^{R,\mathcal{A}}$
via (\ref{eq:jointexit}) and then applying (\ref{eq:matrixexp}). The former
approach would involve repeated multiplications of square matrices of size
$\left|  {\dom A_{i}}\right|  $, for $i=1,\ldots,k$, resulting in a
computational complexity of $\bigO(\sum_{i=1}^{k}n_{i}{\left|  {\dom
A_{i}}\right|  }^{\nu})$, where $n_{i}$ is the number of iterations needed to
compute the sum (\ref{eq:matrixexp}) to machine precision, and the complexity
of multiplying two $n\times n$ matrices is $\bigO(n^{\nu})$.\footnote{$\nu=3$ for
the standard method and $\nu=\log_{2}7$ for Strassen's and related methods.}
The latter would involve multiplying square matrices of size $\prod_{i=1}%
^{k}\left|  {\dom
A_{i}}\right|  $, resulting in the considerably worse complexity of
$\bigO(n(\prod_{i=1}^{k}{\left|  {\dom A_{i}}\right|  )}^{\nu})$, where $n$ is the
number of iterations needed to obtain the desired precision.

\subsection{Simplified modification models}

We next introduce several possible simplifications of the general Markov chain
case. To do so, we start by differentiating numeric domains from non-numeric
domains. Certain database attributes $A\in\mathcal{A}$, such as prices and
order quantities, represent numbers, and numeric operations such as
addition are meaningful for these attributes. For such attributes, one can
easily define a distance function between two attribute values, as we shall
see below. We call the domains $\dom A$ of such attributes \emph{numeric
domains}, and denote the set of all attributes with numeric domains by
$\mathcal{N}\subset\mathcal{B}$. All other attributes and domains are
considered \emph{non-numeric}.\footnote{Distance metrics can also be defined
for complex data types such as images. We leave the handling of such cases to
further research.} It is worth noting that not all numeric data necessarily
constitute a numeric domain. Consider, for example, a customer relation $R$
whose primary key is a customer number. Although the customer number consists
of numeric symbols, it is essentially an arbitrary identification string for
which arithmetic operations like addition and subtraction are not
intrinsically meaningful for the database application. We consider such
attributes to be non-numeric.

\subsubsection{Domain lumping}

\label{sec:lumping}To make our data modification model more computationally
tractable, it may be appropriate, in many cases, to simplify the Markov chain
state space for an attribute $A$ so that it is much smaller than $\dom A$.
Suppose, for example, that $A$ is a 64-character string representing a street
address. Restricting to 96 printable characters, $A$ may assume on the order
of $96^{64}\approx10^{126}$ possible values. It is obviously unnecessary,
inappropriate, and intractable to work with a Markov chain with such an
astronomical number of states.

One possible remedy for such situations is referred to as \emph{lumping }in
the Markov chain literature~\cite{KEMENY60}. In our terminology, suppose we
can partition $\dom A$ into a collection of sets ${\{V\}}_{V\in\mathcal{V}}$
with the property that $\left|  {\mathcal{V}}\right|  \ll\left|  {\dom
A}\right|  $ and
\[
\forall\;U,V\in\mathcal{V},\;\forall\;u,u^{\prime}\in U\quad\sum_{v\in
V}q_{u,v}^{R,A}=\sum_{v\in V}q_{u^{\prime},v}^{R,A}.
\]
Then, one can model the transitions between the ``lumps'' $V\in\mathcal{V}$ as
a much smaller Markov chain whose set of states is $\mathcal{V}$, with the
transition rate from $U\in\mathcal{V}$ to $V\in\mathcal{V}$ being given by the
common value of $\sum_{v\in V}q_{u,v}^{R,A}$, $u\in U$. If we are interested
only in which lump the attribute is in, rather than its precise value, this
smaller chain will suffice. Using lumping, the complexity of the computation
is directly dependent on the number of lumps. We now give a few simple examples:

\begin{example}
[Lumping into a binary domain]\label{ex:binarylump}Consider the
street address example just discussed. Fortunately, if an address has changed
since time $s$, the database user is unlikely to be concerned with how
different it is from the address at time $s$, but simply whether it is
different. Thus, instead of modeling the full domain $\dom A$, we can
represent the domain via the simple binary set $\{0,1\}$, where $0$
indicates that the address has not changed since time $s$, and $1$ indicates
that it has. We assume that the exit rates $q_{v,r.A(s)}^{R,A}$ from all other
addresses $v\in\dom A$ back to the original value $r.A(s)$ all have the
identical value $\theta^{\prime}$. In this case, one has $P_{0,1}%
^{R,A}=P_{1,0}^{R,A}=1$, and the behavior of the attribute is fully captured
by the exit rates $\ell_{0}^{R,A}=q_{0,1}^{R,A}$ and 
$\ell_{1}^{R,A}=q_{1,0}^{R,A}$.  We will abbreviate
these quantities by $\theta$ and $\theta^{\prime}$, respectively.

Using standard results for a two-state continuous-time Markov chain
\cite[Section VI.3.3]{TK94}, we conclude that
\begin{align}
P_{0,0}^{R,A}(s,f) &  =
\frac{\theta^{\prime}+\theta e^{-{(\theta+\theta^{\prime})\Gamma_{R,A}(s,f)}}}
{\theta+\theta^{\prime}}\label{binary:0to0} \\
P_{0,1}^{R,A}(s,f) &  =
\frac{\theta-\theta e^{-{(\theta+\theta^{\prime})\Gamma_{R,A}(s,f)}}}
{\theta+\theta^{\prime}}.  \label{binary:0to1}%
\end{align}
\hspace*{\fill}$\Box$
\end{example}


\begin{example}[Web crawling]
As an even simpler special case, consider a Web crawler
(\emph{e.g.}, \cite{PINKERTON94,HEYDON99,CHO00}). Such a crawler needs to
visit Web pages upon change to re-process their content, possibly for the use
of a search engine. Recalling Example \ref{ex:binarylump}, one
may define a boolean attribute \texttt{Modified} in a relation that collects
information on Web pages. \texttt{Modified} is set to \texttt{True} once the
page has changed, and back to \texttt{False} once the Web crawler has visit
the page. Therefore, once a page has been modified to \texttt{True}, it cannot
be modified back to \texttt{False} before the next visit of the Web
crawler. 
In the analysis of 
Example \ref{ex:binarylump}, one can set $\theta^{\prime}=0$,
resulting in $P_{0,0}^{A,R}(s,f)=e^{-{\theta\Gamma_{R,A}(s,f)}}$ and $P_{0,1}%
^{A,R}(s,f)=1-e^{-{\theta\Gamma_{R,A}(s,f)}}$. \hspace*{\fill}$\Box$
\end{example}

\subsubsection{Random walks}
\label{sec:randomwalks} 
Like large non-numeric domains, many numeric domains may
also be cumbersome to model directly via Markov chain techniques. For example,
a 32-bit integer attribute can, in theory, take $2^{32}\approx4\times10^{9}$
distinct values, and it would be virtually impossible to directly form, much
less exponentiate, a full transition rate matrix for a Markov chain of this size.

Fortunately, it is likely that such attributes will have ``structured'' value
transition patterns that can be modeled, or at least closely approximated, in
a tractable way. As an example, we consider here a random walk model for
numeric attributes.

In this case, we still suppose that the attribute $A$ is modified only at
transition event times that are distributed as described above. Letting
$t_{i}$ denote the time of transition event $i$, with $t_{0}=s$, we suppose
that at transition event $i$, the value of attribute $A$ is modified according
to
\[
r.A(t_{i})=r.A(t_{i-1})+\Delta A_{i},
\]
where $\Delta A_{i}$ is a random variable. We suppose that the random
variables $\left\{  \Delta A_{i}\right\}  $ are IID, that is, they are
independent and share a common distribution with mean $\delta$ and variance
$\sigma^{2}$. Defining
\[
\Delta A(s,f)=\!\!\!\sum_{i:t_{i}\in(s,f]}\!\!\!\!\Delta A_{i},
\]
we obtain that $\{\Delta A(s,f),f\geq s\}$ is a nonhomogeneous compound
Poisson process, and $r.A(f)=r.A(s)+\Delta A(s,f)$. From standard results for
compound Poisson processes, we then obtain for each tuple $r\in R(s)$ that
$\expecop\!\left[  {r.A(f)}\right]  =r.A(s)+\Gamma_{R,A}(s,f)\delta$.

It should be stressed that such a model must ultimately be only an
approximation, since a random walk model of this kind would, strictly
speaking, require an infinite number of possible states, while $\dom A$ is
necessarily finite for any real database. However, we still expect it to be
accurate and useful in many situations, such as when $r.A(s)$ and
$\expecop\!\left[  {r.A(f)}\right]  $ are both far from largest and smallest
possible values in $\dom A$.

\subsubsection{Content-independent overwrites}

Consider the simple case in which $P_{u,v}^{R,A}(s,f)$ is independent of $u$
once a transition event has occurred. Let $\mathcal{A}\subseteq\mathcal{C}(R)$
be a set of attributes $A$ with identical $\gamma_{R,A}$ functions, and let
$\Gamma_{R,\mathcal{A}}(s,t)=\Gamma_{R,A}(s,t)$ for any $A\in\mathcal{A}$. We
define a probability distribution $\omega_{R,\mathcal{A}}$ over $\dom
\mathcal{A}$, and assume that at each transition event, a new value for
$\mathcal{A}$ is selected at random from this distribution, without regard to
the prior value of $r.\mathcal{A}$. It is thus possible that a transition
event will leave $r.\mathcal{A}$ unchanged, since the value selected may be
the same one already stored in $r$. For any tuple $r\in R(s)\cap R(f)$ and
$u\in\dom\mathcal{A}$, we thus compute the probability $P_{u,u}^{R,\mathcal{A}%
}(s,f)$ that the value of $r.\mathcal{A}$ remains unchanged at $u$ at time
$f$ to be
\begin{align*}
P_{u,u}^{R,\mathcal{A}}(s,f)  &  =\probop\!\left\{  {\tau_{u}^{R,\mathcal{A}%
}>f-s}\right\}  +\probop\!\left\{  {\tau_{u}^{R,\mathcal{A}}\leq f-s}\right\}
\omega_{R,\mathcal{A}}(u)\\
&  =e^{-\ell_{u}^{R,\mathcal{A}}\Gamma_{R,\mathcal{A}}(s,f)}+\left(
1-e^{-\ell_{u}^{R,\mathcal{A}}\Gamma_{R,\mathcal{A}}(s,f)}\right)
\omega_{R,\mathcal{A}}(u)\\
&  =e^{-\ell_{u}^{R,\mathcal{A}}\Gamma_{R,\mathcal{A}}(s,f)}\left(
1-\omega_{R,\mathcal{A}}(u)\right)  +\omega_{R,\mathcal{A}}(u)
\end{align*}
For $u,v\in\dom\mathcal{A}$ such that $u\neq v$, we also compute the
probability $P_{u,v}^{R,A}(s,f)$ that $r.\mathcal{A}$ changes from $u$ to $v$
in $[s,f)$ to be
\begin{align*}
P_{u,v}^{R,\mathcal{A}}(s,f)  &  =\probop\!\left\{  {\tau_{u}^{R,\mathcal{A}%
}\leq f-s}\right\}  \omega_{R,\mathcal{A}}(v)\\
&  =\left(  1-e^{-\ell_{u}^{R,\mathcal{A}}\Gamma_{R,\mathcal{A}}(s,f)}\right)
\omega_{R,\mathcal{A}}(v).
\end{align*}

Content-independent overwrites are a special case of the Markov chain model
discussed above. To apply the general model formulae when content-independent
updates are present, each $\ell_{v}^{R,A}$ is multiplied by $1-\omega
_{R,A}(v)$ and $P_{u,v}^{R,A}=\omega_{R,A}(v)/(1-\omega_{R,A}(u))$ for all
$u,v\in\dom A$, $u\neq v$.

\subsection{Summary}

In this section we have introduced a general Markov-chain model for data
modification, and discussed three simplified models that allows tractable
computation. Using these models, one can compute, in probabilistic terms, value
histograms at time $f$, given a known initial set of value histograms
at time $s<f$. Such a
model could be useful in query optimization, whenever the
continual gathering of statistics becomes impossible due to either heavy system
loads or structural constraints (\emph{e.g.}, federations of databases with
autonomous DBMSs).

Generally speaking, computing the transition matrix for an attribute $A$
involves repeated multiplications of square matrices of size $\left|  {\dom
A}\right|  $, resulting in a computational complexity of $\bigO(n{\left|  {\dom
A}\right|  }^{\nu})$, where $n$ is the number of iterations needed to compute
the sum (\ref{eq:matrixexp}) to machine precision. While $n$ is usually small,
$\left|  {\dom
A}\right|  $ may be very large, as demonstrated in Section
\ref{sec:lumping} 
and Section \ref{sec:randomwalks}. Methods such as domain lumping would
require $\bigO(nX^{\nu})$ time, where $X\ll\left|  {\dom
A}\right|  $.\footnote{Here, $n$ may also be affected by the change of
domain.} As for random walks and independent updates, both methods no longer
require repeated matrix multiplications, but rather the computation of
$\Gamma_{R,A}(s,f)$. The complexity of calculating $\Gamma_{R,A}(s,f)$ is
similar to that for $\Lambda_{R}(s,f)$ in Section \ref{sec:lambdacomplex}.

\section{Insertion model verification}
\label{sec:verify} 
It is well-known that Poisson processes
model a world where data updates are independent from one another. While in
databases with widely distributed access, \emph{e.g.}, incoming e-mails,
postings to newsgroups, or posting of orders from independent customers, such
an independence assumption seems plausible, we still need to validate the
model against real data. In this section we shall present some initial
experiments as a ``proof of concept.'' These experiments deal only with the
insertion component of the model. Further experiments, including modification
and deletion operations, will be reported in future work.

\begin{figure}[ptb]
\begin{center}
\epsfig{file=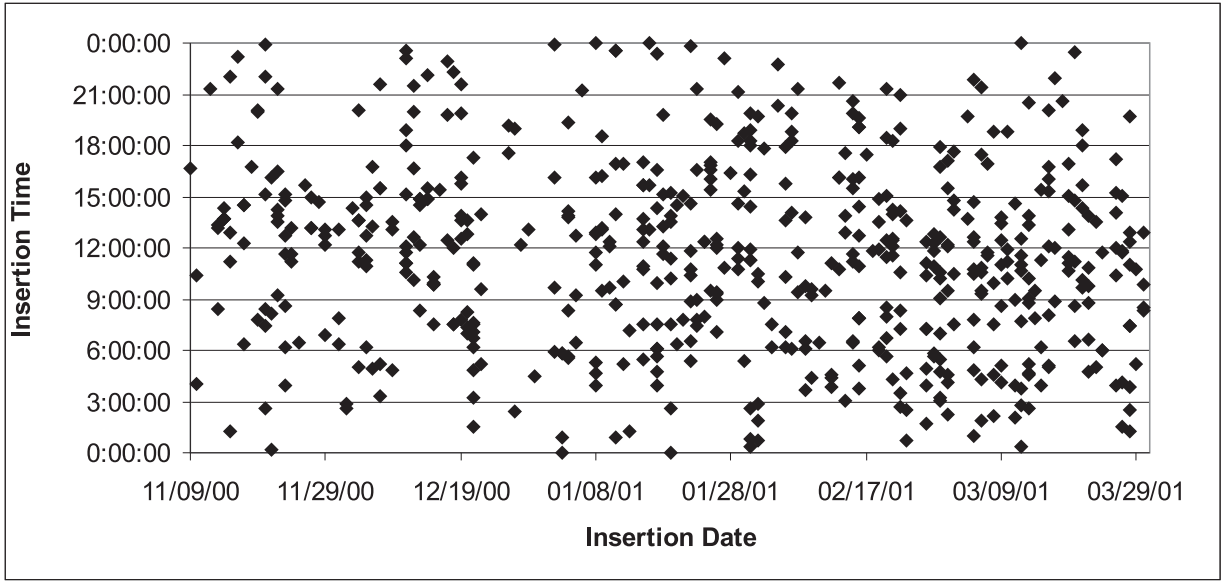,width=6.5in}
\caption{Training data set.}
\label{fig:training}
\end{center}
\end{figure}

Our data set is taken from postings to the DBWORLD electronic bulletin board.
The data were collected over more than seven months and consists of about 750
insertions, from November 9$^{\text{th}}$, 2000 through May 14$^{\text{th}}$,
2001. Figure \ref{fig:training} illustrates a data set with 580
insertions during the interval 
[2000/11/9:00:00:00,~2001/3/31:00:00:00). We used the 
Figure \ref{fig:training} data as a \emph{training set}, 
\emph{i.e.}, it serves as our basis for
parameter estimation. Later, in order to test the model, we applied these
parameters to a separate \emph{testing set} covering the period
[2001/3/31:00:00:00,~2001/5/15:00:00:00). In the experiments described below,
we tried fitting the training data with two insertion-only models, namely a
homogeneous Poisson process and an RPC Poisson process (see Section
\ref{sec:insertion}). For each of these two models, we have applied two
variations, either as a compound or as a non-compound model. In the
experiments described below, we have used the Kolmogorov-Smirnov goodness of
fit test (see for example~\cite[Section 7.7]{HOGG83}). For completeness, we
first overview the principles of this statistical test.

The Kolmogorov-Smirnov test evaluates the likelihood of a \emph{null
hypothesis} that a given sample may
have been drawn from some
hypothesized distribution.  If the null hypothesis is true, and
sample set has indeed been drawn from the
hypothesized distribution, then the empirical cumulative distribution of the
sample should be close to its theoretical counterpart. If the sample
cumulative distribution is too far from the hypothesized distribution at any
point, that suggests that the sample comes from a different distribution.
Formally, suppose that the theoretical distribution is $F(x)$, and we have $n$
sample values $x_{1},...,x_{n}$ in nondecreasing order. We define an empirical
cumulative distribution $F_{n}(x)$ via
\[
F_{n}(x)=\left\{
\begin{array}
[c]{cl}%
0, & \text{if }x<x_{1}\\
\frac{k}{n}, & \text{if }x_{k}\leq x<x_{k+1}\\
1, & \text{if }x>x_{n},
\end{array}
\right.
\]
and then compute $D_{n}=\sup_{k=1,\ldots,n}\{|F_{n}(x_{k})-F(x_{k})|\}$. For
large $n$, given a significance level $\alpha$, the test measures $D_{n}$
against $X(\alpha)/\sqrt{n}$, where $X(\alpha)$ is a factor depending on the
\emph{significance level} $\alpha$ at which we reject the null hypothesis. 
For example, $X(0.05)=1.36$ and $X(0.1)=1.22$.  The value of $\alpha$
is the probability of a ``false negative,'' that is, the chance that
the null hypothesis might be rejected when it is actually true.
Larger values of $\alpha$ make the test harder to pass.

\subsection{Fitting the homogeneous Poisson process}
\label{sec:fithomo}
\begin{figure}[ptb]
\begin{center}
\epsfig{file=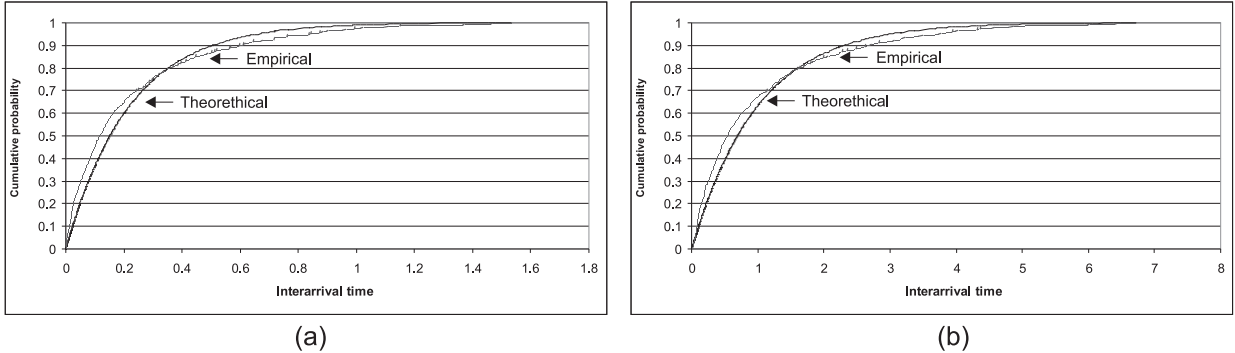, width=6.5in}
\caption{A comparison of a theoretical and empirical distribution functions
for the homogeneous Poisson process model (a) and the compound homogeneous
Poisson process model (b).}%
\label{fig:kshomo}%
\end{center}
\end{figure}

Based on the training set, we computed the parameter for a homogeneous Poisson
process by averaging the 580 interarrival times, an unbiased estimator of the
Poisson process parameter. The average interarrival time was computed to be
5:15:19, and thus $\lambda=4.57$ per day. Figure \ref{fig:kshomo}(a) 
provides a pictorial comparison of the cumulative
distribution functions of the interarrival times with their theoretical
counterpart. We applied the Kolmogorov-Smirnov test to the distribution
of interarrival times, comparing it with an exponential distribution with a
parameter of $\lambda=4.57$. The outcome of the test is $D_{n}=0.106$, which
means we can reject the null hypothesis at any reasonable level of confidence
$\alpha\geq 0.005$ (for $\alpha=0.005$, the rejection threshold is $0.0718$ for
$n=580$). In all likelihood, then, the data are not derived from a homogeneous
Poisson process.

Next, we have applied a compound homogeneous Poisson model.  Our
rationale in this case is that DBWORLD is a moderated list, and the
moderators sometimes work on postings in batches.  
These batches are sometimes posted to the group in tightly-spaced clusters.
For all practical purposes, we
treat each such cluster as a single batch insertion event.
To construct the model, any
two insertions occurring within less than one minute from one another were
considered to be a single event occurring at the insertion
time of the first arrival. For example, on November 14, 2000, we had three
arrivals, one at 13:43:19, and two more at 13:43:23. All three arrivals are
considered to occur at the same insertion arrival event, with an insertion
time of 13:43:19. Using the compound variation, the data set now has 557
insertion events. The revised average interarrival time is now 5:28:20, and
thus $\lambda=4.39$ per day. Figure \ref{fig:kshomo}(b)
provides a pictorial comparison of the cumulative distribution functions of
the interarrival times, assuming a compound model, with their theoretical
counterpart. We have applied the Kolmogorov-Smirnov test to the distribution
of interarrival times, comparing it with an exponential distribution with a
parameter of $\lambda=4.39$. The outcome was somewhat better than before.
$D_{n}=0.094$, which means we can still reject the null hypothesis at any
level of confidence $\alpha\geq 0.005$ (for $\alpha=0.005$, the rejection
threshold is $0.0733$ for $n=557$). Although the compound variant of the model
fits the data better, it is still not statistically plausible.

\subsection{Fitting the RPC Poisson process}
\label{sec:rpcfit}%
\begin{table}[tbp] \centering
\begin{tabular}[c]{|l|l|l|l|}\hline
& \textbf{Workdays} & \textbf{Saturday} & \textbf{Sunday}\\\hline\hline
$\lbrack0\text{:}00,3\text{:}00)$ & \multicolumn{1}{|c|}{$2.40$} &
\multicolumn{1}{|c|}{} & \multicolumn{1}{|c|}{}\\\cline{1-2}%
$\lbrack3\text{:}00,6\text{:}00)$ & \multicolumn{1}{|c|}{$5.96$} &
\multicolumn{1}{|c|}{} & \multicolumn{1}{|c|}{}\\\cline{1-2}%
$\lbrack6\text{:}00,9\text{:}00)$ & \multicolumn{1}{|c|}{$6.04$} &
\multicolumn{1}{|c|}{} & \multicolumn{1}{|c|}{}\\\cline{1-2}%
& & \multicolumn{1}{|c|}{$1.50$} & \multicolumn{1}{|c|}{$1.15$}\\
$\lbrack9\text{:}00,18\text{:}00)$ & \multicolumn{1}{|c|}{$7.50$} &
\multicolumn{1}{|c|}{} & \multicolumn{1}{|c|}{}\\
& & \multicolumn{1}{|c|}{}& \multicolumn{1}{|c|}{}\\\cline{1-2}%
$\lbrack18\text{:}00,21\text{:}00)$ & \multicolumn{1}{|c|}{$3.03$} &
\multicolumn{1}{|c|}{} & \multicolumn{1}{|c|}{}\\\cline{1-2}
$\lbrack21\text{:}00,24\text{:}00)$ & \multicolumn{1}{|c|}{$2.41$} &
\multicolumn{1}{|c|}{} & \multicolumn{1}{|c|}{}\\\hline
\end{tabular}
\caption{Average $\lambda$ levels for the recurrent piecewise-constant 
Poisson model.}
\label{tab:rpclambdas}
\end{table}

Next, we tried fitting the data to an RPC model.
Examining the data, we chose a cycle of one week. 
Within each week, we used the same pattern for each weekday, with one
interval for work hours (9:00-18:00), plus five additional three-hour
intervals for ``off hours''.  
We treated Saturday and Sunday each as one long interval.
Table \ref{tab:rpclambdas}
shows the arrival rate parameters for each segment
of the RPC Poisson model, calculated in much the same manner as the
for the homogeneous Poisson model. 

The specific methodology for structuring the RPC Poisson model is
beyond the scope of this paper and can range from \emph{ad hoc}
``look and feel'' crafting (as practiced here)
to more established formal processes for
statistically segmenting, filtering, and aggregating intervals
\cite{STOUMBOS97,STOUMBOS2002}. It is worth noting, however, that from
experimenting with different methods, we have found that the model is not
sensitive to slight changes in the interval definitions.  Also, the model we
selected has only $8$ segments, and thus only $8$ parameters, so there
is little danger of ``overfitting'' the training data set, which has over
$500$ observations.

Next, we attempted to statistically validate the RPC model. To this end, we use
the following lemma:

\begin{lemma}
\label{lem:udistrib}
Given a nonhomogeneous Poisson process with arrival intensity
$\lambda(t)$, the random variable $U_{s}=\int
_{s}^{s+L_{R,s}}\lambda(t)\hspace{0.15em}dt$ is of the distribution
$\expdistrib(1)$.
\end{lemma}

\begin{proof}
Let $f_{s}(t)=\Lambda(s,s+t)$, which is a monotonically nondecreasing
function. From Lemma \ref{lem:interarrival}, $\probop
\!\{{L_{R,s}<t\}=}1-e^{f_{s}(t)}$ for all $t\geq0$. We have $U_{s}=f_{s}(L_{R,s}%
)$. By applying the monotonic function $f_{s}$ to both sides of the inequality
$L_{R,s}<t$, one has that $\probop\!\{f_{s}({L_{R,s})<f_{s}(t)\}}=\probop
\!\{{L_{R,s}<t\}}=1-e^{f_{s}(t)}$ for all $t\geq0$. Substituting in the
definitions of $U_{s}$ and $u=f_{s}(t)$, one then obtains $\probop
\!\{U{_{s}<u\}}=1-e^{-u}$ for all $u\geq0$, and therefore $U_{s}%
\sim\expdistrib(1)$.
\end{proof}

Thus, given an instantaneous arrival rate $\lambda(t)$, and a sequence of
observed arrival events $\{t_{n}\}_{n=0}^{N}$, we compute the set of values
$u_{n}=\int_{t_{n-1}}^{t_{n}}\lambda(t)\hspace{0.15em}dt$, $n=1,\ldots,N,$ and
perform a Kolmogorov-Smirnov test of them versus the unit exponential
distribution. 

\begin{figure}[tb]
\begin{center}
\epsfig{file=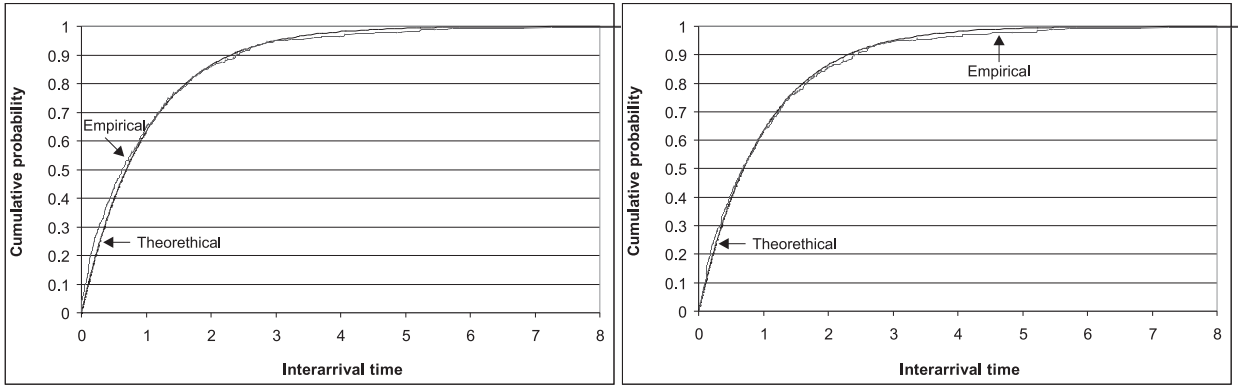,width=6.5in}
\caption{A comparison of a theoretical and empirical distribution functions of
$U$ for the RPC Poisson model (a) and the compound RPC Poisson model (b).}
\label{fig:ksrpc}%
\end{center}
\end{figure}

Figure \ref{fig:ksrpc}(a) provides a
comparison of the theoretical and empirical cumulative distribution of the
random variable $U$. We applied the Kolmogorov-Smirnov test to $U$,
comparing it with an exponential distribution with $\lambda=1$, based on Lemma
\ref{lem:udistrib}. The outcome of the test is $D_{n}=0.080$,
which is better than either homogeneous model, but is still rejected
at any reasonable level of significance
(recall that for $\alpha=0.005$, the rejection threshold is again $0.0718$ for $n=580$). 

Finally, we evaluated a compound version of the RPC model, combining
successive postings separated by less than one minute.  We kept the
same segmentation as in Table \ref{tab:rpclambdas}, but recalculated the
arrival intensities in each segment, as shown in
Table~\ref{tab:compoundrpclambdas}. 

\begin{table}[tbp] \centering
\begin{tabular}[c]{|l|l|l|l|}
\hline
& \textbf{Workdays} & \textbf{Saturday} & \textbf{Sunday}\\\hline\hline
$\lbrack0\text{:}00,3\text{:}00)$ & \multicolumn{1}{|c|}{$2.40$} &
\multicolumn{1}{|c|}{} & \multicolumn{1}{|c|}{}\\\cline{1-2}%
$\lbrack3\text{:}00,6\text{:}00)$ & \multicolumn{1}{|c|}{$5.96$} &
\multicolumn{1}{|c|}{} & \multicolumn{1}{|c|}{}\\\cline{1-2}%
$\lbrack6\text{:}00,9\text{:}00)$ & \multicolumn{1}{|c|}{$5.59$} &
\multicolumn{1}{|c|}{$$} & \multicolumn{1}{|c|}{$$}\\\cline{1-2}%
& &
\multicolumn{1}{|c|}{$1.45$} & \multicolumn{1}{|c|}{$1.15$}\\
$\lbrack9\text{:}00,18\text{:}00)$ & \multicolumn{1}{|c|}{$7.11$} &
\multicolumn{1}{|c|}{} & \multicolumn{1}{|c|}{}\\
& & \multicolumn{1}{|c|}{} & \multicolumn{1}{|c|}{}\\\cline{1-2}%
$\lbrack18\text{:}00,21\text{:}00)$ & \multicolumn{1}{|c|}{$3.03$} &
\multicolumn{1}{|c|}{} & \multicolumn{1}{|c|}{}\\\cline{1-2}
$\lbrack21\text{:}00,24\text{:}00)$ & \multicolumn{1}{|c|}{$2.33$} &
\multicolumn{1}{|c|}{} & \multicolumn{1}{|c|}{}\\\hline
\end{tabular}
\caption{Average $\lambda$ levels for the compound RPC Poisson model.}
\label{tab:compoundrpclambdas}
\end{table}

Next we recalculated the sample of the random variable $U$ for the
compound RPC Poisson model, and applied the Kolmogorov-Smirnov
test. In this case, we have $D_{n}=0.050$, which cannot be rejected at
any reasonable confidence level through $\alpha=0.10$ (for $\alpha=0.10$, the
rejection threshold is $0.0517$ for $n=557$).  Figure
\ref{fig:ksrpc}(b) 
shows the
theoretical and empirical distributions of $U$ in this case.

\begin{figure}[tb]
\begin{center}
\epsfig{file=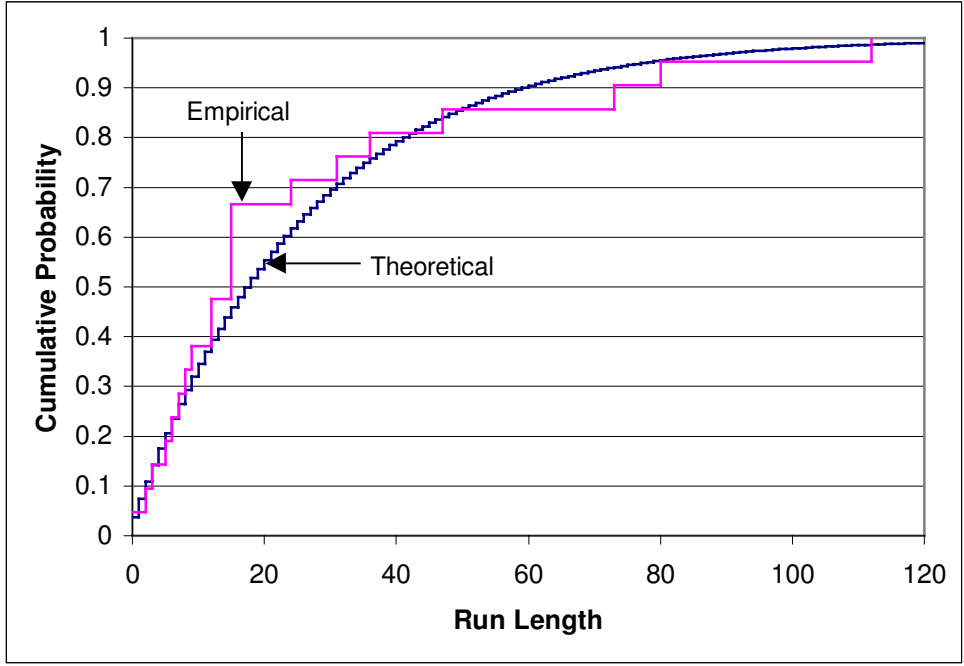}
\caption{Empirical and theoretical distributions for number of
single arrivals between multiple arrivals, compound RCP Poisson model.}
\label{fig:runs}
\end{center}
\end{figure}

As a final confirmation of the applicability of the compound RPC
Poisson model, we attempted to validate the assumption that the number
of postings in successive insertion events are independent and
identically distributed (IID).  In the sample, 536 insertion events
were of size 1, 19 were of size 2, and 2 were of size 3.  Thus, we
approximate the random variable $\Delta^+_R$ as having a $536/557
\approx .962$ probability of being 1, a $19/557 \approx .034$
probability of being 2, and a $2/557 \approx .004$ probability of
being 3.  Validating that the observed insertion batch sizes
$\Delta^+_{R,i}$ appear to be independently drawn from this
distribution is somewhat delicate, since they nearly always take the
value 1.  To compensate, we performed our test on the 
\emph{runs} in the sample, that is, the number of consecutive insertion
events of size 1 between insertions of size 2 or 3.  Our sample
contains 21 runs, ranging from 0 to 112.  If the insertion
batch sizes $\{\Delta^+_{R,i}\}$ are independent with the distribution
$\Delta^+_R$, then the length of a run should be a geometric random
variable with parameter $536/557\approx .962$.  We tested this
hypothesis via a Kolmogorov-Smirnov test, as shown in
Figure~\ref{fig:runs}.  The $D_n$ statistic is $0.207$, which is 
within the $\alpha=0.1$ acceptance level for a sample of size $n=21$
(although the divergence of the theoretical and empirical curves in
Figure~\ref{fig:runs} is more visually pronounced than in the prior figures, it
should be remembered that the sample is far smaller).
Thus, the assumption that the insertion batch sizes
$\{\Delta^+_{R,i}\}$ are IID is plausible.

\begin{table}[tbp] \centering
\begin{tabular}[c]{|l|c|c|}\hline
\textbf{Model} & $D_{n}$ & \textbf{Rejection level}\\\hline\hline
Homogeneous & {$0.106$} & {$<0.005$}\\\hline
Homogeneous+compound & {$0.094$} &
{$<0.005$}\\\hline
RPC & {$0.080$} & {$<0.005$}\\\hline
RPC+compound & {$0.050$} & {$>0.100$}\\\hline
\end{tabular}
\caption{Goodness of fit of the four models.}
\label{tab:goodfit}
\end{table}

Table \ref{tab:goodfit} compares the goodness-of-fit of
the four models to the test data. For each of the models, we have specified
the KS test result ($D_{n}$) and the level at which one can reject the null
hypothesis. The higher the level of confidence is, the better the fit is. The
RPC compound Poisson model models best the data set, accepting the null
hypothesis at any level up to $0.1$ (which practically means that the model
can fit to the data well). The main conclusion from these experiments is that
the simple model of homogeneous Poisson process is limited to the modeling of
a restricted class of applications (one of which was suggested in
\cite{CHO00}). Therefore, there is a need for a more elaborate model, as
suggested in this paper, to capture a broader range of update behaviors. A
nonhomogeneous model consisting of just 8 segments per week, 
as we have constructed, seems to model the arrivals significantly
better than the homogeneous approach.

\section{Content evolution cost model}

\label{costmodel} We now develop a cost model suitable for
transcription-scheduling applications such as those described in
Example~\ref{ex:replica}. The question is how often to generate a remote
replica of a relation $R$. We have suggested one such policy in Example
\ref{ex:fat}. In this section, we shall introduce two more policies and
show an empirical comparison based on the data introduced in Section
\ref{sec:verify}.

A transcription policy aims to minimize the combined cost of
\emph{transcription cost} and \emph{obsolescence cost}~\cite{GAL99c}. The
former includes the cost of connecting to a network and the cost of
transcribing the data, and may depend on the time at which the transcription
is performed (\emph{e.g.}, as a function of network congestion), and the
length of connection needed to perform the transcription. The obsolescence
cost captures the cost of using obsolescent data, and is 
basically a function of the
amount of time that has passed since the last transcription.

In what follows, let the set $\{b_{i},e_{i}\}_{i=1}^{\infty}$ represents an
infinite sequence of connectivity periods between a client and a server.
During session $i$, the client data is synchronized with the state of the
server at time $b_{i}$, the information becoming available at the client at
time $e_{i}$. At the next session, beginning at time $b_{i+1}$, the client is
updated with all the information arriving at the server during the interval
$(b_{i},b_{i+1}]$, which becomes usable at time $e_{i+1}$, and so forth. We
define $b_{0}=e_{0}=0$, and require that $0<b_{1}\leq e_{1}<b_{2}\leq
e_{2}<\ldots$.

Let $C_{R,\text{u}}(s,f)$ denote the cost of performing a transcription of $R$
starting at time $f$, given that the last update was started at time $s$. Let
$C_{R,\text{o}}(s,f)$, to be described in more detail later, denote the
obsolescence cost through time $f$ attributable to tuples inserted into $R$ at
the server during the time interval $(s,f]$. Then the total cost $C_{R}(t)$
through time $t$ is
\begin{equation}
C_{R}(t)=\sum_{i:b_{i}\leq t}\!\!
\Big(  
\alpha C_{R,\text{u}}(b_{i-1},b_{i})
+(1-\alpha)C_{R,\text{o}}(b_{i-1},b_{i})
\Big) + (1-\alpha)C_{R,\text{o}}(b_{i^*(t)},t), 
\label{costformula}
\end{equation}
where $i^*(t)=\max\left\{i\;\big|\;b_{i}\leq t\right\}$ and
$\alpha$ serves as the ratio of importance a user puts on the
transcription cost versus the obsolescence cost. Traditionally, $\alpha=0$,
and therefore $C_{R}(t)$ is minimized for $C_{R,\text{o}}(b_{i-1},b_{i})=0$,
$\forall b_{i}<t$, allowing the use of current data only. In this section we
shall look into another, more realistic approach, where data currency is
sacrificed (up to a level defined by the user through $\alpha$) for the sake
of reducing the transcription cost. Ideally, one would want to choose the
sequence $\{b_{i},e_{i}\}_{i=1}^{\infty}$ of connectivity periods, subject to
any constraints on their durations $e_{i}-b_{i}$, to minimize $C_{R}(t)$ over
some time horizon $t$. One may also consider the asymptotic problem of
minimizing the average cost over time, $\lim_{t\rightarrow\infty}C_{R}(t)/t$.
We note that the presence of $\alpha$ is not strictly required, as its effects
could be subsumed into the definitions of the $C_{R,\text{u}}$ and
$C_{R,\text{o}}$ functions, especially if both are expressed in natural
monetary units. However, we retain $\alpha$ in order to demonstrate some of
the parametric properties of our model.

In general, modeling transcription and obsolescence costs may be difficult and
application-dependent. They may be difficult to quantify and difficult to
convert to a common set of units, such as dollars or seconds. Some subjective
estimation may be needed, especially for the obsolescence costs. However, we
maintain that, rather than avoiding the subject altogether, it is best to try
construct these cost models and then use them, perhaps parametrically, to
evaluate transcription policies. Any transcription policy implicitly makes
some trade-off between consuming network resources and incurring
obsolescence, so it is
best to try quantify the trade-off and see if a better policy exists. In
particular, one should try to avoid policies that are clearly \emph{dominated}%
, meaning that there is another policy with the same or lower transcription
cost, and strictly lower obsolescence, or \emph{vice versa}. Below, for
purposes of illustration, we will give one simple, plausible way in which the
cost functions may be constructed; alternatives are left to future research.

\subsection{Transcription costing example}
In determining the transcription cost, one may use existing research into
costs of distributed query execution strategies. Typically, (\emph{e.g.},
\cite{LOHMAN85}) the transcription time can be computed as some function of
the CPU and I/O time for writing the new tuples onto the client and the cost
of transmitting the tuples over a network. There is also some fixed setup time
to establish the connection, which can be substantial. For purposes of
example, suppose that
\begin{align*}
C_{R,\text{u}}(s,f)  &  =c+\beta\cdot\left(  X_{R}(s,f)+Y_{R}^{+}(s,f)+\left|
R(s)\right|  -Y_{R}(s,f)\right) \\
&  = c+\beta\cdot\left(  X_{R}(s,f)+\left|  R(s)\right|  -Y_{R}^{-}%
(s,f)\right)
\end{align*}
Here, $c\geq0$ denotes the fixed setup cost, $\beta\geq0$, $X_{R}(s,f)$
denotes the number of tuples inserted during the interval $(s,f]$ that survive
through time $f$, $Y_{R}^{+}(s,f)$ is the number of tuples that
survive but are
modified, by time $f$, and $\left|  R(s)\right|  -Y_{R}(s,f)$ is the number of
deleted tuples. For the latter, it may suffice to transmit only the 
primary key of each deleted tuple, incurring a unit cost of less than
$\beta$. For sake of simplicity, however, we use the same cost factor
$\beta$ for deletion, insertion, and modification. We note that, under
this assumption,
\[
\sum_{i:b_{i}\leq t}C_{R,\text{u}}(b_{i-1},b_{i})=n(t)c+\beta\left|
R(s)\right|  +\beta\sum_{i:b_{i}\leq t}\left(  X_{R}(b_{i-1},b_{i})-Y_{R}%
^{-}(b_{i-1},b_{i})\right)  ,
\]
where $n(t)$ is the number of transcriptions in the interval $[0,t]$. For the
special case that there are no deletions or modifications, 
$\beta\left|  R(s)\right|
+\beta\left(  X_{R}(s,f)-Y_{R}^{-}(s,f)\right)  =\beta B(s,f)$ and
\[
\sum_{i:b_{i}\leq t}C_{R,\text{u}}(b_{i-1},b_{i})=n(T)c+\beta B(0,b_{i^{\ast
}(T)}).
\]
For large $t$, one would expect the $\beta B(0,b_{i^{\ast}(t)})$ term to be
roughly comparable across most reasonable polices, whereas the $n(t)c$ term
may vary widely for any value of $t$. It is worth noting that $c$ and $\beta$
could be generalized to vary with time or other factors. 
For example, due to network congestion, certain
times of day may have higher unit transcription costs than others. 
Also, transcribing via airline-seat telephone costs substantially more than
connecting via a cellular phone. For simplicity, we have refrained from
discussing such variations in the transcription cost.

\subsection{Obsolescence costing example}
We next turn our attention to the obsolescence cost, which is clearly a
function of the update time of tuples and the time they were transcribed to
the client. Intuitively, the shorter the time between the update of a tuple
and its transcription to the client, the better off the client would be. As a
basis for the obsolescence cost, we suggest a criterion that takes into
account user preferences, as well as the content evolution parameters. For any
relation $R$, times $s<f$, and tuple $r\in R(s)\cup R(f)$, let $b(r)$ and
$d(r)$ denote the time $r$ was inserted into and deleted from $R$,
respectively. We let $\iota_{r}(s,f)$ be some function denoting the
contribution of tuple $r$ to the obsolescence cost over $(s,f]$; we will give
some more specific example forms of this function later. We then make the
following definition:

\begin{definition}
The total \emph{obsolescence cost} of a relation $R$ over the time interval
$(s,f]$ (annotated \emph{$C_{R,\text{o}}(s,f)$}) is defined to be
\emph{$C_{R,\text{o}}(s,f)\triangleq\sum_{r\in R(s)\cup R(f)}\iota
_{r}(s,f)\!\!.$}\hspace*{\fill}$\Box$
\end{definition}

Our principal concern is with the \emph{expected} 
obsolescence cost, that is, the
expected value of $C_{R,\text{o}}(s,f)$,
\[
\expecop\!\left[  C_{R,\text{o}}(s,f)\right]  = \expecop\!\left[
\sum_{r\in R(s)\cup R(f)}\!\!\!\!\!\!\!\!\iota_{r}(s,f)\right]  .
\]
To compute $\expecop\!\left[  C_{R,\text{o}}(s,f)\right]  $, we note that
\[
\expecop\!\left[  C_{R,\text{o}}(s,f)\right]  =
\expecop\!\left[  \sum_{r\in R(s)\cap R(f)}\!\!\!\!\!\!\!\!
\iota_{r}(s,f)\right]  
+\expecop\!\left[  {\sum_{r\in R(s)\backslash R(f)}\!\!\!\!\!\!\!\!
\iota_{r}(s,f)}\right]  
+\expecop\!\left[{\sum_{r\in R(f)\backslash R(s)}\!\!\!\!\!\!\!\!
\iota_{r}(s,f)}\right]  .
\]
The three terms in the last expression represent potentially modified tuples,
deleted tuples, and inserted tuples, respectively. We denote these three terms
by $\hat{\iota}_{R}^{\modification}(s,f)$, $\hat{\iota}_{R}^{\deletion
}(s,f)$, and $\hat{\iota}_{R}^{\medspace\insertion}(s,f)$, respectively,
whence
\[
\expecop\!\left[  C_{R,\text{o}}(s,f)\right]  =\hat{\iota}_{R}^{\modification
}(s,f)+\hat{\iota}_{R}^{\deletion
}(s,f)+\hat{\iota}_{R}^{\medspace\insertion}(s,f).
\]

\subsection{Obsolescence for insertions}
\label{sec:insertobs}
We will now consider a specific
metric for computing the obsolescence stemming from insertions in $(s,f]$, as
follows:
\begin{equation}
\iota_{r}^{\insertion}(s,f)=\left\{
\begin{array}
[c]{ll}%
g^{\insertion}(s,f,b(r)) & s<b(r)\leq f<d(r)\\
0 & \text{otherwise},%
\end{array}
\right.  \label{iota}%
\end{equation}
where $g^{\insertion}(s,f,t)$ is some application-dependent function representing the
level of importance a user assigns, over the interval $(s,f]$,
to a tuple arriving at a time $t$. For
example, in an e-mail transcription application, a user may attach greater
importance to messages arriving during official work hours, and a lesser
measure of importance to non-work hours (since no one expects her to be
available at those times). Thus, one might define
\begin{equation}
g^{\insertion}(s,f,t)=\int_{t}^{f}a(\tau)\hspace{0.15em}d\tau,\quad\text{where }%
a(\tau)=\left\{
\begin{array}
[c]{ll}%
a_{1}, & \text{if }\tau\text{ is during work hours}\\
a_{2}, & \text{if }\tau\text{ is after hours,}%
\end{array}
\right.  \label{importanceformula}%
\end{equation}
and $a_{1}\geq a_{2}$. For $a_{1}=a_{2}=1$, $g^{\insertion}(s,f,t)$ 
takes a form resembling the age of a local element in~\cite{CHO00}.
More complex forms of $g^{\insertion}(s,f,t)$ are certainly possible.  In this
simple case, we refer to $a_1/a_2$ as the \emph{preference ratio}.

Using the properties of nonhomogeneous Poisson processes, we calculate
\begin{align*}
\hat{\iota}_{R}^{\medspace\insertion}(s,f) &  =\expecop\!\left[  {\sum_{r\in
R(f)\backslash R(s)}\!\!\!\!\!\!\!\!}\iota_{r}(s,f)\right]  \\
&  =\expecop\!\left[  {X}_{R}{(s,f)}\right]  \cdot\expecop\!\left[
{f(s,f,b(r))\;\big|\;s<b(r)\leq f<d(r)}\right]  \\
&  =\widetilde{\Lambda}_{R}(s,f)\expecop\!\left[  {\Delta_{R}^{+}}\right]
\int_{s}^{f}\frac{\lambda_{R}(t^{\prime})}{\widetilde{\Lambda}_{R}%
(s,f)}g^{\insertion}(s,f,t^{\prime})dt^{\prime}\\
&  =\expecop\!\left[  {\Delta}_{R}^{+}\right]  \int_{s}^{f}\lambda
_{R}(t^{\prime})g^{\insertion}(s,f,t^{\prime})dt^{\prime}.
\end{align*}%

\begin{example}
[Transcription policies using the expected obsolescence cost]
Consider the insertion-only data set of Section~\ref{sec:verify}.
Figure
\ref{fig:ttimes} compares two transcription
policies for the week 
$[2001/4/2\mathrm{:}0\mathrm{:}00,2001/4/8\mathrm{:}0\mathrm{:}00)$. The
transcription policy in Figure \ref{fig:ttimes}(a) 
(referred to below as the \emph{uniform synchronization
point} --- USP --- policy) was suggested in
\cite{CHO00}. According to this policy, the intervals $(s,f]$ are always of
the same size. The decision regarding the interval size $f-s$ may be either
arbitrary (\emph{e.g.}, once a day) or may depend on $\lambda$, the Poisson
model parameter (in which case a homogeneous Poisson process is implicitly
assumed). The policy may be expressed as $f=s+{M}/{\lambda}$ for some
multiplier $M>0$. According to this policy with $M=1$ (as suggested in
\cite{CHO00}), and $\lambda=4.57$ per day as computed from the training data.
Therefore, one would refresh the database every 5:15:19. 
Figure \ref{fig:ttimes}(a) shows
the transcription times resulting from the 
USP policy.

\begin{figure}[tbp]
\begin{center}
\epsfig{file=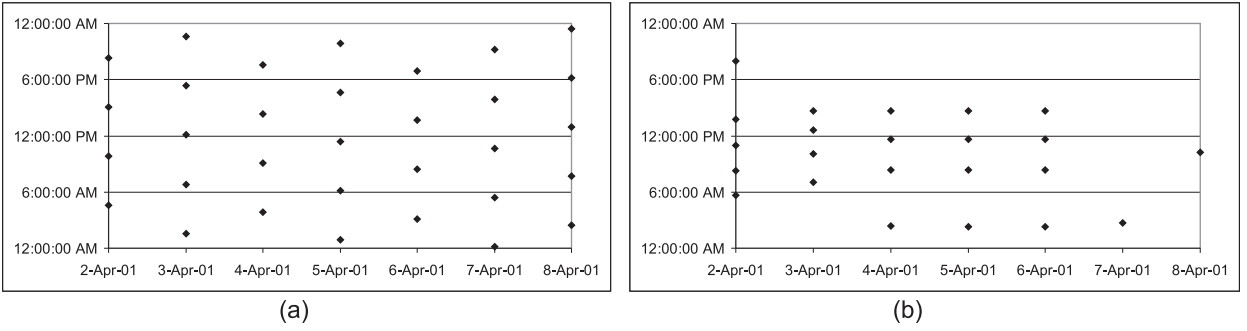,width=6.5in}
\caption{Transcription times for the USP and RPC/threshold policies.}%
\label{fig:ttimes}%
\end{center}
\end{figure}

Consider now another transcription policy, dubbed the
\emph{threshold} policy. With this policy, given that the last connection
started at time $s$, we transcribe at time $f$ if the expected obsolescence
cost from insertions ($\hat{\iota}_{R}^{\medspace\insertion}(s,f)$) exceeds
$\Pi$, where $\Pi$ is a threshold that measures the user's tolerance to
obsolescent data. In comparing the two policies, one can compute $\Pi$, given
$M$, as follows. Consider the homogeneous case where $\expecop\!\left[
{\Delta}_{R}^{+}\right]  =1$ and $\lambda_{R}(t)=\lambda_{R}$ for all $t$.
Assume further that $a_{1}=a_{2}=1$ for all $t$. In this case,
\[
\hat{\iota}_{R}^{\medspace\insertion}(s,f)=\int_{s}^{f}\lambda_{R}%
\cdot(f-t)dt=\lambda_{R}\int_{s}^{f}(f-t)dt\newline =\frac{1}{2}\lambda
_{R}\cdot(f-s)^{2}%
\]
Setting $f=s+M/\lambda_{R}$ and $\Pi=\hat{\iota}_{R}^{\medspace\insertion
}(s,f)$, one has that
\[
\Pi=({1}/{2})\lambda_{R}\cdot(f-s)^{2}=({1}/{2})\lambda_{R}\cdot\left(
{M}/{\lambda}_{R}\right)  ^{2}={M^{2}}/{2\lambda}_{R}.
\]
Figure \ref{fig:ttimes}(b) shows the
transcription times using the RPC arrival model (see Section 
\ref{sec:fithomo}) and the threshold policy with
$\Pi=0.109$ (obtained by setting $M=1$ and $\lambda=4.57$ per day, and letting
$\Pi={M^{2}}/{2\lambda}$). It is worth noting that transcriptions are more
frequent when the $\lambda$ intensity is higher and less frequent whenever the
arrival rate is expected to be more sluggish.

\begin{figure}[tbp]
\begin{center}
\epsfig{file=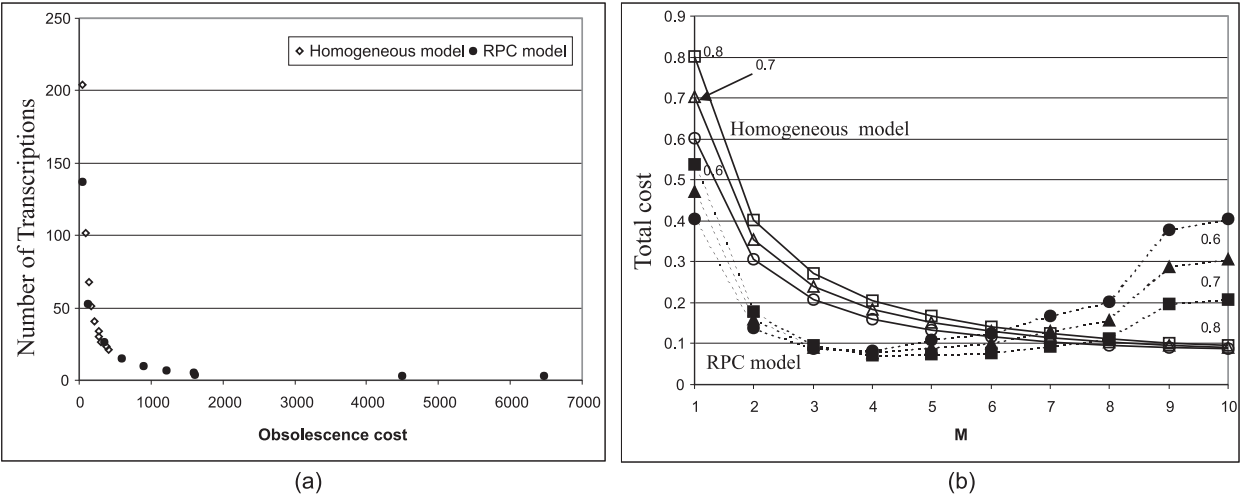,width=6.5in}
\caption{Threshold policy, homogeneous vs. RPC.}%
\label{fig:thpolicy}%
\end{center}
\end{figure}

We have performed experiments comparing the performance of the threshold
policy for the homogeneous Poisson model (equivalent to the USP
policy) and the
RPC Poisson model. Figure \ref{fig:thpolicy} shows
representative results, with costs computed over the testing set. 
Figure \ref{fig:thpolicy}(a)
displays the obsolescence cost and the number of transcriptions for various
$M$ values, with a preference ratio $a_2/a_1=4$. For all $M$ values, there is
no dominant model. For example, for $M=1$, the RPC model has a slightly higher
obsolescence cost (43.02 versus 42.35, a 1.6\% increase) 
and a significantly lower
number of transcriptions (137 versus 204, a 32.8\% decrease).

Figure \ref{fig:thpolicy}(b) provides a comparison of
combined normalized obsolescence and transcription costs for both insertion
models and $\alpha\in\{0.6,0.7,0.8\}$ (still assuming a 4:1 preference ratio).
Solid lines represent results related with the homogeneous Poisson model,
while dotted lines represent results related with the RPC Poisson model.
Generally speaking, the RPC model performs better for small $M$ values
($M\leq 7$), 
while the homogeneous model performs better for the largest $M$ values
($M \geq 8$). \hspace*{\fill}$\Box$
\end{example}

\begin{figure}[tbp]
\begin{center}
\epsfig{file=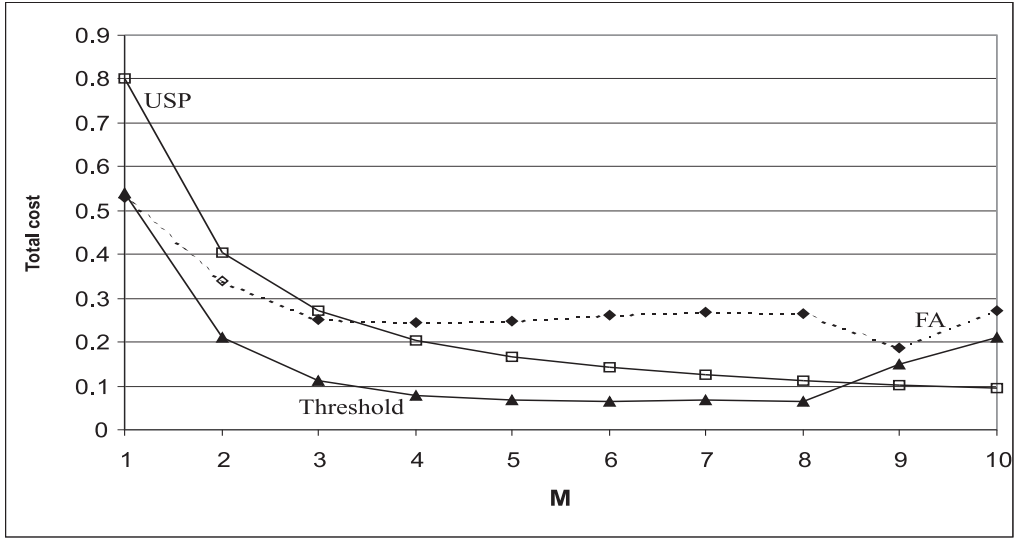,width=5.4in,height=2.9in}
\caption{Comparison of three policies, for a 4:1 preference ratio and
$\alpha=0.5$.}
\label{fig:3policies}
\end{center}
\end{figure}

\begin{example}
[Comparison of USP, threshold and FA policies]
Once again with the data from Section~\ref{sec:verify}, 
we consider one more transcription policy, the first alteration (FA) policy
derived from the analysis
of Example \ref{ex:fat}.  Since there are no deletions or
modifications, $Z_R(s,f)$ simplifies to $\Lambda_R(s,f)$.  We choose
$\pi$ in the FA policy 
to be a function of $M$ such that the transcription intervals
agree with the USP policy in the case of the homogeneous model.
Figure \ref{fig:3policies} 
compares the performance of all three transcription policies: 
USP, threshold, and FA,
for a 4:1 preference ratio and
$\alpha=0.8$, using the testing data set to compute the costs. 
For $M=1$, the threshold policy and the FA policy perform similarly, where the FA policy performs slightly better than the Threshold policy. Both policies outperform the USP policy. 
The threshold policy is best for $M\in\{2 \dots 8\}$. For all $M>8$, the USP
policy is best. The best policy for this choice of $a_{1}/a_{2}$ and $\alpha$
is threshold with $M=6$, followed closely by FA with $M\in\{5,6\}$. We have
conducted our experiments with various $\alpha$ values and our conclusion is
that the Threshold model is preferred over the USP model for larger $\alpha$, that is, the more the user is willing to sacrifice currency for the sake of reducing transcription cost.\hspace*{\fill}$\Box$
\end{example}

\begin{figure}[tbp]
\begin{center}
\epsfig{file=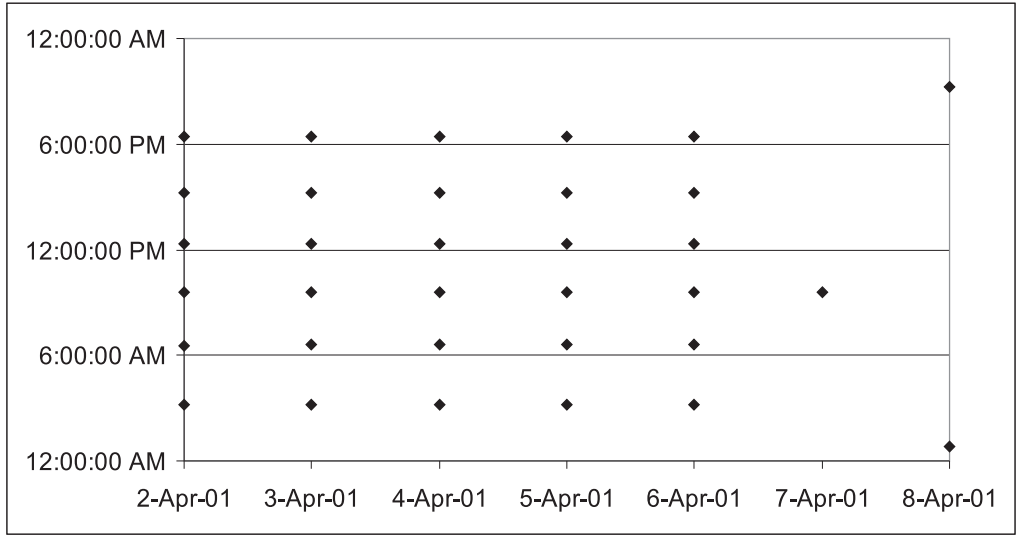,width=5.4in,height=2.9in}
\caption{Transcription schedule based on the first alteration policy.}
\label{fig:fapolicy}
\end{center}
\end{figure}

\subsection{Obsolescence for deletions}
In a similar manner to Section \ref{sec:insertobs}, 
we will consider the following
metric for computing the obsolescence stemming from deletions in $(s,f]$. We
compute $\iota_{r}^{\deletion}(s,f)$ via
\begin{equation}
\iota_{r}^{\deletion}(s,f)=\left\{
\begin{array}
[c]{ll}%
g^{\deletion}(s,f,d(r)) & b(r)\leq s<d(r)\leq f\\
0 & \text{otherwise}%
\end{array}
\right.
\end{equation}
where $g^{\deletion}(s,f,t)$ is some application-dependent
function,  possibly similar to $g^{\insertion}(s,f,t)$ above.

Using the properties of nonhomogeneous Poisson processes, we calculate
\begin{align*}
\hat{\iota}_{R}^{\deletion}(s,f)  &  =\expecop\!\left[  {\sum_{r\in
R(s)\backslash R(f)}\!\!\!\!\!\!\!\!}\iota_{r}(s,f)\right] \\
&  =\left(  \left|  R(s)\right|  -\expecop\!\left[  {Y}_{R}{(s,f)}\right]
\right)  \expecop\!\left[  g^{\deletion}(s,f,d(r)){\;\big|\;b(r)\leq s<d(r)\leq
f}\right] \\
&  =\left(  \left|  R(s)\right|  -p_{R}(s,f)\left|  {R(s)}\right|  \right)
\expecop\!\left[  g^{\deletion}(s,f,d(r)){\;\big|\;b(r)\leq s<d(r)\leq f}\right] \\
&  =\left|  R(s)\right|  \left(  1-p_{R}(s,f)\right)  
\expecop\!\left[  g^{\deletion}(s,f,d(r)){\;\big|\;b(r)\leq s<d(r)\leq f}\right]
\end{align*}
In the case $\langle R,S\rangle$ has fixed multiplicity for all $S\in S(R)$,
$p_{R}(s,f)=\exp(-\widetilde{M}_{R}(s,f))$, where $\widetilde{M}_{R}%
(s,f)=\int_{s}^{f}\tilde{\mu}_{R}(t)\hspace{0.15em}dt$ and $\tilde{\mu}%
_{R}(t)=\!\!\sum_{S\in S(R)}w(R,S)\mu_{S}(t)$. Therefore,%
\begin{align*}
\hat{\iota}_{R,A}^{\deletion}(s,f)  &  =\left|  R(s)\right|  \left(
1-\exp(-\widetilde{M}_{R}(s,f))\right)  \int_{s}^{f}\frac{\tilde{\mu}%
_{R}(t^{\prime})}{\widetilde{M}_{R}(s,f)}g^{\deletion}(s,f,t^{\prime})dt^{\prime}\\
&  =\left|  R(s)\right|  \left(  \frac{1-\exp(-\widetilde{M}_{R}%
(s,f))}{\widetilde{M}_{R}(s,f)}\right)  \int_{s}^{f}\tilde{\mu}%
_{R}(t^{\prime})\cdot g^{\deletion}(s,f,t^{\prime})dt^{\prime}%
\end{align*}

\subsection{Obsolescence for modification}
We now consider obsolescence costs relating to modifications.
While, in some applications, a user may be
primarily concerned with how many tuples were modified during $[s,f)$, 
we believe that a more general, attribute-based framework is warranted
here, taking into account exactly how each tuple was changed.
Therefore, we define $\iota_{r,A}(s,f)$ to be some function denoting the
contribution of attribute $A\in\mathcal{A}(R)$ in tuple $r$ to the
obsolescence cost over $(s,f]$ and assume that%
\[
\iota_{r}(s,f)=\!\!\!\sum_{A\in\mathcal{A}(R)}\!\!\!\iota_{r,A}(s,f)
\]
Therefore,
\begin{align*}
\hat{\iota}_{R}^{\modification}(s,f)  & 
=\expecop\!\left[  \sum_{r\in R(s)\cap R(f)}\!\!\!\!\!\iota_{r}(s,f)\right]  \\
& =\expecop\!\left[  \sum_{A\in\mathcal{A}(R)}\;\sum_{r\in R(s)\cap R(f)}
\!\!\!\!\!\!\!\iota_{r,A}(s,f)\right]  \\
& =\sum_{A\in\mathcal{A}(R)}\!\!\!\hat{\iota}_{R,A}^{\modification}(s,f)
\end{align*}
where $\hat{\iota}_{R,A}^{\modification}(s,f)$ 
is the expected obsolescence cost due to modifications to $A$ during
$(s,f]$.  Assuming that attributes not in $\mathcal{C}(R)$ incur zero
modification cost, the last sum may be taken over $\mathcal{C}(R)$
instead of $\mathcal{A}(R)$.

We start the section by introducing the notion of distance metric and provide
two models of $\iota_{r,A}(s,f)$, for numeric and non-numeric domains. We then
provide an explicit description of $\hat{\iota}_{R,A}^{\modification
}$, based on distance metrics. 

\subsubsection{General distance metrics}
Let $c_{u,v}^{R,A}$, where $u,v\in\dom A$ denote the
elements of a matrix of costs for an attribute $A$. We declare that if
$r.A(s)=u$ and $r.A(f)=v$, then $\iota_{r,A}(s,f)=c_{u,v}^{R,A}$, or
equivalently,
\[
\iota_{r,A}(s,f)=c_{r.A(s),r.A(f)}^{R,A}.
\]
Consequently, we require that $c_{u,u}^{R,A}=0$ for all $u\in\dom A$, so that
an unchanged attribute field yields a cost of zero.

\paragraph{A squared-error metric for numeric domains:}
For numeric domains, that
is, $A\in\mathcal{N}$, we propose a squared-error metric, as is standard in
statistical regression models. In this case, we let
\[
\iota_{r,A}(s,f)=c_{r.A(s),r.A(f)}^{R,A}=k_{R,A}(s){\left(
r.A(f)-r.A(s)\right)  }^{2},
\]
where $k_{R,A}(s)$ is a user-specified
scaling factor. A typical choice for the scaling
factor would be the reciprocal ${1}/\left(  {\varop_{r\in R(s)}\!\left[
{r.A(s)}\right]  }\right)  $ of the 
variance of attribute $A$ in $R$ at time $s$,
\begin{align*}
\varop_{r\in R(s)}\!\left[  {r.A(s)}\right]   &  =\expecop_{r\in
R(s)}\!\left[  {{\left(  r.A(s)-\expecop_{r\in R(s)}\!\left[  {r.A(s)}\right]
\right)  }^{2}}\right] \\
&  =\expecop_{r\in R(s)}\!\left[  {{r.A(s)}^{2}}\right]  -{\expecop_{r\in
R(s)}\!\left[  {r.A(s)}\right]  }^{2}\\
&  =\frac{1}{\left|  {R(s)}\right|  }\left(  \,\sum_{v\in\dom A}%
\!\!\!v^{2}\hat{R}_{A,v}(s)\right)  -{\left(  \frac{1}{\left|  {R(s)}\right|
}\sum_{v\in\dom A}\!\!\!v\hat{R}_{A,v}(s)\right)  }^{2}.
\end{align*}
Other choices for the scaling factor $k_{R,A}(s)$ are also possible. In any
case, we may calculate the expected alteration cost for attribute $A$ in tuple
$r$ via
\begin{align}
\expecop\!\left[  \iota_{r,A}(s,f)\right]   &  =\expecop\!\left[
{k_{R,A}(s){\left(  r.A(f)-r.A(s)\right)  }^{2}}\right] \nonumber\\
&  =k_{R,A}(s)\expecop\!\left[  {{r.A(f)}^{2}-2\,r.A(f)r.A(s)+{r.A(s)}^{2}%
}\right] \nonumber\\
&  =k_{R,A}(s)\left(  \expecop\!\left[  {{r.A(f)}^{2}}\right]
-2\,r.A(s)\expecop\!\left[  {r.A(f)}\right]  +{r.A(s)}^{2}\right)  .
\label{eq:gennumexpec}%
\end{align}

\paragraph{A general metric for non-numeric domains:}
\label{nonnummetric}For non-numeric domains, it may not be possible or
meaningful to compute the difference of $r.A(s)$ and $r.A(f)$. In such cases,
we shall use a general cost matrix ${[}${$c_{u,v}^{R,A}$}${]}_{u,v\in\dom A}$
and compute
\begin{align*}
\expecop\!\left[  \iota_{r,A}(s,f)\right]   &  =\sum_{v\in\dom A}\!\!\left(
P_{r.A(s),v}^{R,A}(s,f)\right)  \left(  c_{r.A(s),v}^{R,A}\right) \\
&  =\sum_{
\genfrac{}{}{0pt}{1}{v\in\dom A}{v\neq r.A(s)}%
}\!\!\left(  P_{r.A(s),v}^{R,A}(s,f)\right)  \left(  c_{r.A(s),v}%
^{R,A}\right)  .
\end{align*}

For domains that have no particular structure, a typical choice might be
$c_{u,v}^{R,A}=1$ whenever $u\neq v$. In this case, the expected cost
calculation simplifies to
\begin{align*}
\expecop\!\left[  \iota_{r,A}(s,f)\right]   &  =\probop\!\left\{  {r.A(f)\neq
r.A(s)}\right\} \\
&  =1-P_{r.A(s),r.A(s)}^{A,R}(s,f).
\end{align*}

We are now ready to consider the calculation of $\hat{\iota}_{R,A}%
^{\modification}(s,f)$.

\subsubsection{The expected modification cost}
We next consider computing the
expected modification cost $\hat{\iota}_{R,A}^{\modification}(s,f)$. To do so,
we partition the tuples $r$ in $R(s)\cap R(f)$ according to their initial
value $r.A(s)$ of the attribute $A$. Consider the subset $R_{A,u}(s)\cap R(f)$
of all $r\in R(s)\cap R(f)$ that have $r.A(s)=u$. Since all such tuples are
indistinguishable from the point of view of the modification process for
$(R,A)$, their $\iota_{r,A}(s,f)$ random variables will be identically
distributed. The number of tuples $r\in R(s)$ with $r.A(s)=u$ is, by
definition, $\hat{R}_{A,u}(s)$. The number $\left|  {R_{A,u}(s)\cap
R(f)}\right|  $ that are also in $r.A(f)$ is a random variable whose
expectation, by the independence of the deletion and modification processes,
must be $p_{R}(s,f)\hat{R}_{A,u}(s)$. Using standard results for sums of
random numbers of IID random variables, we conclude that
\begin{align*}
\hat{\iota}_{R,A}^{\modification}(s,f)  &  =\expecop\!\left[  {\sum_{r\in
R(s)\cap R(f)}\!\!\!\!\!\!\!\!}\iota_{r,A}(s,f)\right] \\
&  =\!\!\!\!\sum_{u\in\dom A}\!\!\!\left(  p_{R}(s,f)\hat{R}_{A,u}(s)\right)
\expecop\!\left[  \iota_{r,A}(s,f){\;\big|\;r.A(s)\!=\!u}\right] \\
&  =p_{R}(s,f)\!\!\!\!\!\!\sum_{%
\genfrac{}{}{0pt}{1}{u\in\dom A}{\hat{R}_{A,u}(s)>0}%
}\!\!\!\!\!\!\!\hat{R}_{A,u}(s)\expecop\!\left[  \iota_{r,A}(s,f){\;\big
|\;r.A(s)\!=\!u}\right] \\
&  =p_{R}(s,f)\!\!\!\!\!\!\!\sum_{%
\genfrac{}{}{0pt}{1}{u\in\dom A}{\hat{R}_{A,u}(s)>0}%
}\!\!\!\!\!\!\hat{R}_{A,u}(s)\hat{\iota}_{R,A,u}^{\modification}(s,f),
\end{align*}
where we define $\hat{\iota}_{R,A,u}^{\modification}(s,f)=\expecop\!\left[
\iota_{r,A}(s,f){\;\big|\;r.A(s)\!=\!u}\right]  $. We now address the
calculation of the $\hat{\iota}_{R,A,u}^{\modification}(s,f)$.

For a non-numeric domain, we have from Section \ref{nonnummetric} that
\[
\hat{\iota}_{R,A,u}^{\modification}(s,f)=\!\!\!\sum_{v\in\dom A}%
\!\!\!\!\!\left(  P_{u,v}^{R,A}(s,f)\right)  \left(  c_{u,v}^{R,A}\right)  ,
\]
and in the simple case of $c_{u,v}^{R,A}=1$ whenever $u\neq v$,
\[
\hat{\iota}_{R,A,u}^{\modification}(s,f)=1-P_{u,u}^{R,A}(s,f).
\]
In any case, $P_{u,v}^{R,A}(s,f)$ and $P_{u,u}^{R,A}(s,f)$ may be computed
using the results of Section \ref{sec:modif}.

For a numeric domain, we have from (\ref{eq:gennumexpec}) that
\begin{align*}
\hat{\iota}_{R,A,u}^{\modification}(s,f)  &  =k_{R,A}(s)\left(  \expecop
\!\left[  {{\left(  r.A(f)\right)  }^{2}\;\big|\;r.A(s)\!=\!u}\right]
-2\,u\expecop\!\left[  {r.A(f)\;\big|\;r.A(s)\!=\!u}\right]  +u^{2}\right) \\
&  =k_{R,A}(s)\left(  \left(  \sum_{v\in\dom A}\!\!\!(v^{2}-2uv)P_{u,v}%
^{R,A}(s,f)\right)  +u^{2}\right)  .
\end{align*}

In cases where a random walk approximation applies, however, the situation
simplifies considerably, as demonstrated in the following proposition.

\begin{proposition}
When a random walk model with mean $\delta$ and variance $\sigma^2$
accurately describes 
modifications to a numeric attribute $A$, $\hat{\iota
}_{R,A,u}^{\modification}(s,f)\approx 
k_{R,A}(s)\,\Gamma_{R,A}(s,f)\left(  \sigma
^{2}+2\,\Gamma_{R,A}(s,f)\delta^{2}\right)  .$
\end{proposition}

\begin{proof}
\noindent In this case, we note that the random variable $r.A(f)-r.A(s)$ is
identical to $\Delta A(s,f)$ 
(using the notation of section \ref{sec:randomwalks}%
), and is independent of $r.A(s)$. The number $N$ of modification events in
$(s,f]$ has a Poisson distribution with mean $\Gamma_{R,A}(s,f)$, and hence
variance {$\Gamma_{R,A}(s,f)$}$^{2}$. Therefore we have, for any $u\in\dom
A$,
\begin{align*}
\hat{\iota}_{R,A,u}^{\modification}(s,f) &  \approx k_{R,A}(s)\expecop\!\left[
{{\left(  \Delta A(s,f)\right)  }^{2}}\right]  \\
&  =k_{R,A}(s)\left(  \varop\!\left[  {\Delta A(s,f)}\right]  +{\expecop
\!\left[  {\Delta A(s,f)}\right]  }^{2}\right)  \\
&  =k_{R,A}(s)\left(  \expecop\!\left[  {N}\right]  \sigma^{2}+\delta
^{2}\varop\!\left[  {N}\right]  +{\expecop\!\left[  {N}\right]  }^{2}%
\delta^{2}\right)  \\
&  =k_{R,A}(s)\,\Gamma_{R,A}(s,f)\left(  \sigma^{2}+2\,\Gamma_{R,A}%
(s,f)\delta^{2}\right)  .
\end{align*}
\hspace*{\fill}
\end{proof}

\subsection{Example: the use of the cost model in Web crawling}

\label{sec:exwebcraw} The following example concludes the introduction of the cost function. We show
how, by using the cost model, one can generate an optimal transcription policy
for Web crawling.

\begin{example}
[Web Monitoring]WebSQL \cite{MENDELZON97} is a Web monitoring tool which uses
a virtual database schema to query the structural properties of Web documents.
The database schema consists of two relations, \texttt{Document} with six
attributes, namely \texttt{url, title, text, type, length, }and \texttt{modif}%
, and \texttt{Anchor} with four attributes, namely \texttt{base, label, href},
and \texttt{context}. Each tuple in \texttt{Anchor} indicates that document
\texttt{base} contains a link to document \texttt{href}. Consider the
following query (taken from \texttt{http://www.cs.toronto.edu/\symbol{126}%
websql/}), which identifies locally reachable documents that contain some
hyperlink to a compressed Postscript File:

\vspace{2ex} \texttt{SELECT d.url, d.modif }

\texttt{FROM Document d SUCH THAT ``http://www.OtherDoc.html'' -%
$>$%
-%
$>$%
* d, }

\texttt{Anchor a SUCH THAT base = d }

\texttt{WHERE filename(a.href) CONTAINS ``.ps.Z''; }

\vspace{2ex}
(We refrain from dwelling
on the language specification;he interested reader is referred to the cited
Web site.)
Assume that the cost of performing the query at time $t$ is $\sum_{d\in
D(t)}\psi_{d}$, where $D(t)$ represents the set of scanned documents and
$\psi_{d}$ is a random variable representing the size of document $d$ in
bytes. Assuming the $\{\psi_{d}\}$ are IID, the expected cost of performing
the query at time $t$ is thus
\[
\expecop\left[  \sum_{d\in D(t)}\!\!\psi_{d}\right]  =\expecop[\card{D(t)}%
]\expecop[\psi],
\]
where $\psi$ is a generic random variable distributed like the $\{\psi_{d}\}$.

A modification to a document is identified using changes to the \texttt{modif}
attribute of the Document relation. For brevity in what follows, we let
$R=\text{\texttt{Document}}$ and $A=\text{\texttt{modif}}$. We assign the
following costs to changes in $A$:

\begin{itemize}
\item $g^{\deletion}(s,f,t) = 0$ for all $s<t<f$, 
that is, the user has no interest in being
notified of deleted documents.

\item For all $s<t<f$ and $u,v\in\dom A$, $u\neq v$, 
$c^{R,A}_{u,v}=g^{\insertion}(s,f,t)=\expecop[\psi]$, 
where $c_{{R,A}}^{\modification}$ is the cost for
a modified document. For all other attribute $A^{\prime}\neq A$, 
$c^{R,A^{\prime}}_{u,v}=0$ for all $u,v\in\dom A^{\prime}$.
\end{itemize}

Suppose that a query was performed at time $s$, scanning the set of documents
$D(s)$, and returning the set of documents $B(s)$, where $\left|
{B(s)}\right|  \leq\left|  {D(s)}\right|  $. A user is interested in
refreshing the query result without overloading system resources, thus
balancing the cost of refreshing the query results against the cost of using
partial or obsolescent data. This trade-off can be captured by the following
policy: refresh the query at time $f$, after performing it at time $s$ iff
\[
\expecop\!\!\left[  \sum_{d\in D(f)}\!\!\psi_{d}\right]  \;<\;\expecop
\!\left[  C_{R,\mathrm{o}}(s,f)\right]
\]
Thus, an equivalent conditions is
\begin{align*}
\expecop[\card{D(f)}]\expecop[\psi]\; &  <\;\sum_{A^{\prime}\in\mathcal{A}%
(R)}\hat{\iota}_{R,A^{\prime}}^{\modification}(s,f)+\hat{\iota}_{R}%
^{\deletion}(s,f)+\hat{\iota}_{R}^{\medspace\insertion}(s,f)\\
&  =\hat{\iota}_{R,A}^{\modification}(s,f)+\hat{\iota}_{R}^{\medspace
\insertion}(s,f),
\end{align*}
or
\begin{align*}
&  \left(  p_{R}(s,f)\left|  {D(s)}\right|  +\widetilde{\Lambda}%
_{R}(s,f)\expecop\!\left[  {\Delta}_{R}^{+}\right]  \right)  \expecop[\psi
]\;\\
&  <\left(  p_{R}(s,f)\!\!\!\!\sum_{{u\in\dom A}}\!\!\!\hat{R}_{A,u}%
(s)(1-P_{u,u}^{{A,R}}(s,f))+\widetilde{\Lambda}_{R}(s,f)\expecop\!\left[
{\Delta}_{R}^{+}\right]  p^{\medspace\insertion}\right)  \expecop[\psi],
\end{align*}
where $p^{\medspace\insertion}$ is the probability of a newly-inserted
document being relevant to the query. Cancelling the factor of $\expecop[\psi
]$, another equivalent condition is
\[
p_{R}(s,f)\left|  {D(s)}\right|  +\widetilde{\Lambda}_{R}(s,f)\expecop
\!\left[  {\Delta}_{R}^{+}\right]  \;<\;p_{R}(s,f)\!\!\!\!\sum_{u\in\dom
A}\!\!\!\hat{R}_{A,u}(s)(1-P_{u,u}^{{A,R}}(s,f))+\widetilde{\Lambda}%
_{R}(s,f)\expecop\!\left[  {\Delta}_{R}^{+}\right]  p^{\medspace\insertion},
\]
which is independent of the expected document size. Further assume that
$P_{u,u}^{R{A}}(s,f)=P_{\ast,\ast}^{R,{A}}(s,f)$ is independent of $u$. Then
the refresh condition can be expressed as
\[
p_{R}(s,f)D(s)+\widetilde{\Lambda}_{R}(s,f)\expecop\!\left[  {\Delta}_{R}%
^{+}\right]  \;<\;p_{R}(s,f)\left|  {B(s)}\right|  (1-P_{\ast,\ast}^{{A,R}%
}(s,f))+\widetilde{\Lambda}_{R}(s,f)\expecop\!\left[  {\Delta}_{R}^{+}\right]
p^{\medspace\insertion}%
\]
\hspace*{\fill}$\Box$
\end{example}

\section{Conclusion and topics for future research}
\label{sec:conclusion} 
This paper represents a first step in a new research area,
the stochastic estimation of the consistency of transcribed data over time. We
have also suggested one possible 
technique for assigning a cost to the differences
between two relation extensions, including a means of computing the expected
value of this cost under our stochastic model. We have discussed a number
of potential applications relating managing replicas, query
management, and Web crawling. We have also examined several
strategies for refreshing replicas, although other strategies are
certainly possible.

As an illustration of the low client-side computational demands of the
insertion-only transcription application of our model, 
a Java-based demo, based on the transcription policies described in
\cite{GAL2001} and in this paper, can be accessed at
\texttt{http://rbs.rutgers.edu:6677/}. The demo compares the performance of
various policies using data that exist at a backend mSQL database.

We hope to extend our work to the case where the materialized views are not
simple replications, but are produced by SQL queries that involve selections,
projections, natural joins, and certain types of aggregations. This work will
involve a \emph{propagation algebra} for tracing the base data changes through
a series of relational operators.

This development should make it possible to apply the theory to the management
of more complex queries than presented here. In particular, it will facilitate
a possible approach to managing general materialized view obsolescence on a
query-by-query basis, taking into account current user preferences for query
accuracy and speed. The refresh rate of materialized views in a
periodically-updated data source (such as a data warehouse) can be defined in
terms of data obsolescence, which in turn can be stochastically estimated
using our model for content evolution. In this case, we advocate a three-way
cost model for query optimization~\cite{GAL99c}, in which the query optimizer
evaluates various query plans using three complementary factors, namely
\textit{generation cost}, \textit{transmission cost}, and \textit{obsolescence
cost}. The first two factors take on a conventional interpretation and the
obsolescence cost of a query represents a penalty for basing the query result
on possibly obsolescent materialized views. A query plan using only selection
from a local materialized view, for example, might have lower generation and
transmission costs, but a higher obsolescence cost, than a plan fetching
complete base relations from an extranet and then processing them through a
series of join operations. Our model, when combined with additional techniques
to propagate updates through relational operators, can be used as a basis for
estimating the obsolescence cost. However, developing the propagation algebra
may require some enrichment of our basic model, in particular the 
introduction of dependency between the deletion and modification processes.

We foresee several additional future research directions. One direction
involves the design of efficient algorithms for the numerical computations
required by our model. As it stands so far, the most demanding computations
required are general numerical integration and the matrix exponentiation
formula (\ref{eq:matrixexp}). With regard to integration, we note that, in
practice, the nonhomogeneous Poisson arrival rate functions $\lambda_{R}%
(\cdot)$, $\mu_{R}(\cdot)$, and $\gamma_{R,A}(\cdot)$ will most likely be
chosen to be periodic piecewise low-order polynomials, as suggested in
Section \ref{sec:verify}. In such cases, many of the integrals
needed by the model could be performed in closed form within each time
period.

Further calibration and verification of the models in real situations
is also needed.  So far, we have demonstrated that the insertion model
has plausible applications, but this work needs to be extended to the
deletion and modification models.  Furthermore, the insertion model
may need to be generalized to handle situations where there is
``burstiness'' or autocorrelation in the interarrival times that may require more involved techniques than simply combining very
closely spaced arrivals.

Another future research direction involves applying the model to real-life
settings such as managing a data warehouse. While the model is quite flexible,
a methodology is still needed for structuring Markov chains and estimating the
stochastic model's parameters. Finally, in order to calibrate the cost model,
the issue of measuring user tolerance for data obsolescence 
should be considered.

\section*{Acknowledgments}

We would like to thank Benny Avi-Itzhak, Adi Ben-Israel, David Shanno, Andrzej
Ruszczynski, Ben Melamed, Zachary Stoumbos, and Bob Vanderbei for their help.
Also, we thank Kumaresan Chinnusamy and Shah Mitul for their comparative
research on statistics gathering methods and Connie Lu and Gunjan Modha for
their assistance in designing and implementing the demo.

\bibliographystyle{plain}
\bibliography{bib}
\end{document}